\documentclass{article}

\usepackage[preprint]{arxiv_preprint}

\usepackage[utf8]{inputenc}
\usepackage[T1]{fontenc}
\usepackage{hyperref}
\usepackage{url}
\usepackage{booktabs}
\usepackage{amsmath}
\usepackage{amssymb}
\usepackage{amsfonts}
\usepackage{graphicx}
\usepackage{float}
\usepackage{microtype}
\usepackage{xcolor}
\usepackage{tikz}
\usetikzlibrary{arrows.meta, positioning, shapes.geometric, calc}

\newcommand{\method}{Envisage}
\newcommand{\best}[1]{\textbf{#1}}

\title{Envisage: Diffusion-Based Rhinoplasty Goal Visualization with Mask-Decomposed Evaluation}

\author{%
  Mudit Agarwal\\
  University of Washington\\
  Seattle, WA, USA\\
  \texttt{rajagar@uw.edu}\\
  \And
  Amit D. Bhrany, MD\\
  Department of Otolaryngology--Head and Neck Surgery\\
  University of Washington School of Medicine\\
  Seattle, WA, USA\\
  \texttt{abhrany@uw.edu}\\
}

\begin{document}

\maketitle

\raggedbottom

\begin{abstract}
Localized generative editing needs localized evaluation: full-image identity
metrics are structurally confounded under hard-composited edits. We present
\textbf{\method{}}, a FLUX.1-Fill inpainting reference pipeline for
rhinoplasty goal visualization from a single frontal photograph. The
pipeline combines $8$ rhinoplasty clinical presets (the released framework
also includes $8$ blepharoplasty and $8$ rhytidectomy presets), MediaPipe masks,
and hard-mask compositing. The composite preserves outside-mask pixels by
construction, so full-face identity scores are dominated by copied pixels
rather than by the diffusion backbone. Because full-face identity metrics
cannot grade localized edits, we introduce
\textbf{SurgicalScore}, a mask-decomposed $0$--$1$ protocol scoring edit
direction, edit magnitude, masked LPIPS, realism, and outside-mask preservation;
SS\textsubscript{raw} assigns $0.919$ $[0.918, 0.920]$ to a perfect-predictor
control, anchoring the ceiling.

On $N{=}211$, the paired ArcFace gain (output-to-GT minus input-to-GT) is negative for all methods (\method{} $-0.048$ smallest, vs.\ ICEdit $-0.139$, Kontext $-0.242$, InstructPix2Pix $-0.294$; $p < 10^{-4}$), with external validation on a $457$-pair ASPS/PCA corpus showing a larger negative gap. This decrease suggests the in-mask edit fails to shift identity toward post-op: because copied outside-mask pixels are identical between input and output, the paired difference is likely dominated by the in-mask shift. With SurgicalScore, \method{} achieves the highest score ($0.599$ $[0.579, 0.619]$); \method{} leads on both metrics, but the all-negative ArcFace gap shows that full-face identity is poorly aligned with localized surgical accuracy under hard compositing. Furthermore, a $5$-seed GT-oracle (ground-truth-aware best-of selection, an upper bound rather than a deployable result) reduces the residual ArcFace gap by $73\%$ ($-0.054$ to $-0.015$), with positive output-to-GT gain on $33.9\%$ of cases; SurgicalScore on this subset is also raised from $0.609$ to $0.743$ $[0.725, 0.762]$ ($N{=}207$), indicating candidate-space headroom for a learned ranker. As such, for localized edits under hard compositing, progress should be measured with edit-region fidelity rather than full-face identity metrics. We release \method{}, SurgicalScore, preset definitions, and matched split manifests.
\end{abstract}

\section{Introduction}
\label{sec:intro}

Rhinoplasty accounts for approximately 180,000 procedures annually in the United
States~\citep{asps2024stats}. At this volume, expectation management is the dominant
determinant of patient satisfaction; revision rates of $5$--$15\%$~\citep{neaman2013rhinoplasty,newman2024revision}
reflect, in part, discrepancies between expected and actual outcomes.
Current visualization tools span surgeon-drawn sketches and 2D overlays at one end
to proprietary 3D simulation systems such as Crisalix and Vectra 3D at the other.
The 3D systems require capital investment in stereophotogrammetry hardware and
dedicated capture space, plus per-clinic licensing and trained operators; these
costs restrict access to high-volume metropolitan practices and exclude solo,
rural, and community settings. Sketches and 2D overlays, on the other hand, are surgeon-dependent,
single-view, and not photorealistic. The issue is clear: we need a tool
that is believable, affordable, and faithful to the surgical scope. That
means photorealistic outputs, no specialized capture hardware, and
architectural preservation of non-surgical regions. We are not aware of an existing
single-frontal-photo system that delivers all three.

The gap has two structural roots. First, the methods themselves fall short:
geometric methods~\citep{bookstein1989tps} cannot
synthesize new skin texture, 3D approaches require CT scans or
meshes~\citep{ma2021orthognathic} unavailable in standard clinics, and full-face
diffusion models drift identity across non-surgical regions. Second, the
evaluation is also confounded: full-face evaluation conflates two distinct objectives
(outside-mask identity preservation and inside-mask surgical accuracy) and
reports a single number that mixes them.

LandmarkDiff~\citep{landmarkdiff} established the failure mode. That
system conditioned SD~1.5 on landmark wireframes and regenerated the entire
face before compositing the surgical region back. Decomposed evaluation
showed over $95\%$ of its identity score came from copied pixels: $0.509$
composited, $0.023$ without. The result exposed five architectural
limitations: low-resolution backbone, sparse conditioning, full-face
regeneration, synthetic training data, and unaccounted post-hoc compositing after full-face generation. \method{}
corrects all five.

We invert the formulation. Rather than regenerate the full face and
composite the surgical region back, the pipeline masks the surgical region
from the outset, modifies a monocular depth map at landmark-indexed
Gaussians to encode the target tissue change, runs a pretrained
depth-conditioned diffusion model on the mask only, and composites the
output verbatim with the input outside the mask. Outside-mask identity
preservation is an architectural guarantee, not a backbone assumption: the
only quantity left to measure is inside-mask edit fidelity against paired
postoperative ground truth.

That formulation produces the smallest GT-vs-BL ArcFace gap of any method
evaluated on an $N{=}211$ ASPS+PCA cohort (\method{} $-0.048$ vs.\
ICEdit $-0.139$, Kontext $-0.242$, InstructPix2Pix (IP2P) $-0.294$; $p<10^{-4}$ paired) and
the highest mask-decomposed SurgicalScore ($0.599$ $[0.579, 0.619]$ vs.\
$0.502$ / $0.337$ / $0.229$). The negative ArcFace gap replicates on a
$457$-pair external corpus. No method, including ours, achieves positive
paired ArcFace gain over the unedited input proxy: this is a finding about
the metric, not a failure to report. ArcFace was trained for full-face
verification across changes orders of magnitude larger than rhinoplasty, so
it is poorly aligned with an edit at $5\%$-of-pixels scale.

\paragraph{Our contributions.}
\begin{enumerate}
\item \textbf{Methodological diagnosis.} Full-image identity metrics are
structurally confounded under hard-composited edits. Outside-mask structural
similarity (SSIM)~\citep{wang2004ssim} exceeds
$0.999$ for FLUX backbones wrapped in our composite
(Appendix~\ref{app:composite_full}); the property is rule-driven, not
backbone-driven, and is also observed in our cross-procedure preset
demonstrations (Appendix~\ref{app:cross_proc}).
\item \textbf{SurgicalScore.} A mask-decomposed $0$--$1$ protocol scoring
landmark edit direction, magnitude, masked LPIPS, realism, and outside-mask
preservation. Released with matched HDA paired pre/post split manifests
($N{=}27$ bleph, $N{=}21$ rhino, $N{=}9$ rhytid for cross-procedure use;
the rhinoplasty headline cohort is the larger $N{=}211$ ASPS+PCA pool,
Section~\ref{sec:experiments}). SS\textsubscript{raw} assigns
$0.919$ $[0.918, 0.920]$ to a perfect-predictor control (the no-composite
GT paste), anchoring the ceiling.
\item \textbf{Cross-method, cross-procedure characterization.} No method
exceeds the input proxy under GT ArcFace on $N{=}211$. \method{} produces
the only positive-gap fraction above $5\%$ ($16.1\%$ vs.\ $0.0$ / $4.3$ /
$1.0\%$). SurgicalScore reveals component-level failure modes that ArcFace
alone hides.
\item \textbf{\method{} reference pipeline.} Depth-conditioned FLUX-Fill
inpainting with $8$ rhinoplasty presets (Daniel's taxonomy) plus $8$
blepharoplasty (Tessier) and $8$ rhytidectomy (SMAS) preset definitions
released as infrastructure. Smallest GT-vs-BL ArcFace gap and highest
SurgicalScore of any method evaluated.
\end{enumerate}

\paragraph{Component-level evidence.} On $N{=}211$ ($B{=}10{,}000$, K$=5$
best-of paired permutation), each component is individually
paired-significant: hard-mask composite $\Delta$SS$=+0.034$ ($p{=}0.001$),
anatomy-aware $24$-preset prompt $\Delta$SS$=+0.034$ ($p{=}0.002$),
depth-ControlNet conditioning $\Delta$SS$=+0.023$ ($p{=}0.030$,
Appendix~\ref{app:ablation}). Equipping baselines with the same composite
closes most of the bare-baseline gap; \method{} still leads
ICEdit$+$composite by $+0.058$ ($p<10^{-3}$) and IP2P$+$composite by
$+0.045$ ($p{=}0.008$, Section~\ref{app:composite_ablation}). K$=5$ best-of
upper-bounds the deployed-system contribution; recovering the gain requires
a learned ranker (Appendix~\ref{app:deployable_ranker}: naive rankers do
not).

\section{Related Work}
\label{sec:related}

\paragraph{Surgical prediction.}
\citet{eldaly2022rhinoplastyAIreview} survey simulation and AI methods
for rhinoplasty; the surveyed geometric warp approaches cannot synthesize
new skin texture.
\citet{jung2024rhinoplastygan} train a pix2pix GAN on lateral-profile
silhouettes ($N{=}3{,}030$ pairs) and report $52.5\%$ Visual Turing Test
discrimination. Frontal and internal views are excluded by dataset
construction, leaving the frontal regime open.
\citet{bini2025facial} apply inpainting and depth estimation to facial
reconstruction, a different task from aesthetic rhinoplasty on intact
anatomy. \citet{loorduque2024septorhinoplasty} apply diffusion to
septorhinoplasty visualization but evaluate without paired ground-truth or
identity metrics. \citet{varghaei2025aesthetic} report landmark-based
aesthetic outcome assessment on $1{,}259$ paired patients ($N{=}366$
rhinoplasty), establishing the geometric-plus-identity evaluation regime
our SurgicalScore extends.
PtosisDiffusion~\citep{ptosisdiffusion2024} uses training-free ControlNet
guidance for blepharoptosis; we extend it to depth-conditioned
blepharoplasty, rhinoplasty, and rhytidectomy. Concurrent diffusion work
on orthognathic surgery uses graph-conditioned latent diffusion on lateral
cephalograms~\citep{kim2025gposcnet} or volumetric/multi-view input
unavailable in standard clinics~\citep{ma2021orthognathic}.

\paragraph{Diffusion models and ControlNet.}
Latent diffusion models~\citep{rombach2022latent} generate photorealistic
images from text and spatial conditioning; ControlNet~\citep{zhang2023controlnet}
adds spatial control to frozen backbones via zero-convolution layers. Depth
conditioning maps tissue displacement directly to surface change, a
connection edge or pose conditioning does not provide.
ICEdit~\citep{zhang2025icedit} applies instruction-based editing via
FLUX.1-Fill-dev with a mixture-of-experts LoRA~\citep{hu2022lora} and a diptych formulation; we
evaluate it as a baseline (Section~\ref{sec:results}).

\paragraph{Identity preservation.}
ArcFace~\citep{deng2019arcface} cosine similarity is the standard identity
metric in face research. It conflates identity change in the surgical
region (expected) with identity drift in non-surgical regions (undesired);
our mask-decomposed evaluation separates these terms. Face-recognition
fairness on darker skin tones is documented in the broader
literature~\citep{buolamwini2018gender}, but no prior surgical-prediction
work reports metrics stratified by the Monk Skin Tone
Scale~\citep{monk2023monk}.

\section{Methods}
\label{sec:methods}

\subsection{Pipeline Overview}

We invert the formulation: \method{} masks the surgical region from the
outset rather than regenerating the full face and compositing the surgical
region back. A procedure-specific TPS pre-warp first displaces nasal
landmarks by 2--4~px to encode the intended geometric deformation. Depth
Anything~V2 then estimates a monocular depth map, which landmark-indexed
Gaussian kernels modify to encode the intended tissue displacement.
FLUX.1-Fill-dev, conditioned on the modified depth map via a pretrained
depth ControlNet and a procedure-specific text prompt, regenerates only
the masked surgical region. A hard-mask composite copies all non-surgical
pixels verbatim from the input.

Outside-mask identity preservation is an architectural guarantee, not a
backbone assumption. The remaining modeling problem is inside the mask:
edit fidelity against paired postoperative ground truth.

\begin{figure}[t]
  \centering
  \includegraphics[width=0.55\linewidth]{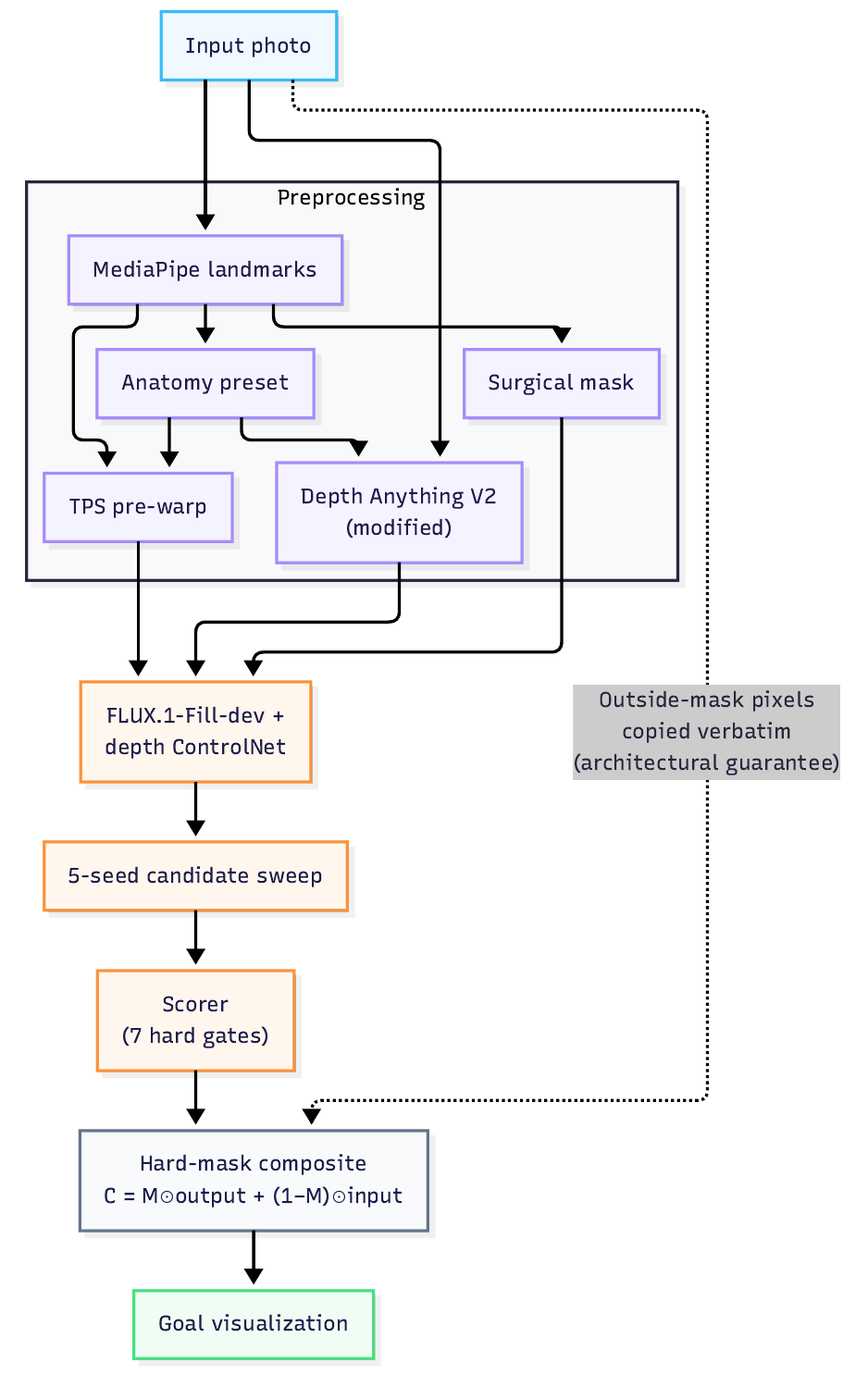}
  \caption{\textbf{Envisage pipeline overview.}
    A preoperative photograph is processed in three stages: (1)~a
    procedure-specific TPS pre-warp displaces nasal landmarks by 2--4~px;
    (2)~Depth Anything~V2 estimates a monocular depth map, which is then
    modified by landmark-indexed Gaussian kernels to encode the intended tissue
    displacement; and (3)~FLUX.1-Fill-dev, conditioned on the modified depth
    map via a pretrained depth ControlNet, regenerates only the masked surgical
    region. A GT-free 7-gate scorer defines candidate validity; headline results
    use a fixed seed, while a GT-oracle best-of-5 over a 5-seed sweep is analyzed
    separately as candidate-space headroom (Appendix~\ref{app:best_of_k}). A
    hard-mask composite pastes all non-surgical pixels verbatim
    from the input, so outside-mask identity preservation is architectural
    rather than a backbone property. We use the \emph{goal visualization}
    framing throughout, not patient-specific outcome prediction.}
  \label{fig:pipeline}
\end{figure}

\subsection{Mask Generation}

For rhinoplasty, 30 MediaPipe FaceLandmarker landmarks (478-point successor of the 468-vertex face mesh~\citep{kartynnik2019mediapipe}) spanning
the nasal dorsum, tip, and alar wings define the convex-hull mask, dilated
25 pixels with an elliptical structuring element and Gaussian-feathered
($\sigma{=}15$ pixels). The preset system (Section~\ref{sec:presets}) supports
8 frontal-detectable rhinoplasty sub-procedures whose masks extend the nasal core
to include the alar base, bridge sidewalls, and supratip region as needed.
The blepharoplasty mask spans 44 upper-lid and brow landmarks; the
rhytidectomy mask spans 36 jaw-contour landmarks. Both follow the same
dilation and Gaussian-feathering protocol as the rhinoplasty mask. Their
preset definitions are released as part of the framework but are not
the primary evaluation target of this paper.

\subsection{TPS Pre-Warp}

For rhinoplasty, bridge sidewall landmarks (indices 193, 245, 188, 174, 217 and
their right-side mirrors) are displaced inward by 3--4 pixels, tip lobule
landmarks inward by 2--3 pixels, and alar landmarks fixed at zero displacement.
The TPS uses \texttt{scipy.interpolate.RBFInterpolator} with a thin-plate
spline kernel and boundary anchor points to prevent deformation outside the
surgical region.

\subsection{Depth Modification}

Monocular depth was estimated with Depth Anything V2 Small~\citep{yang2024depth},
producing $D \in \mathbb{R}^{H \times W}$ normalized to $[0,255]$. Modified
depth is:
\begin{equation}
D'(x,y) = M(x,y) \cdot \bigl[D(x,y) - \textstyle\sum_k \alpha_k \cdot
G_k(x,y; c_k, \sigma_k)\bigr]
+ \bigl[1 - M(x,y)\bigr] \cdot D(x,y),
\label{eq:depth_mod}
\end{equation}
where $M$ denotes the binarized surgical mask used for depth modification (the feathered image-composite mask is a separate, image-domain quantity),
each $G_k$ is a 2D Gaussian at a landmark, and $\alpha_k$ controls
displacement magnitude. For rhinoplasty, two Gaussians are applied: a broad
one at the nasion (landmark~6, $\sigma_x{=}0.05w$, $\sigma_y{=}0.10h$,
$\alpha{=}100$) for dorsal hump reduction, and a tight one at the nasal tip
(landmark~1, $\sigma_x{=}0.015w$, $\sigma_y{=}0.15h$, $\alpha{=}115$) for
the supratip break. Amplitudes $\alpha_k$ are landmark-indexed hand-tuned constants
calibrated once on a held-out HDA case and held fixed for all reported evaluation runs;
evaluation uses 50\% default scaling, applied uniformly without per-cohort retuning. Depth modification does not distinguish
bony from cartilaginous deformation or model surgeon-specific choices such
as spreader-graft placement~\citep{daniel2010mastering}.

\subsection{Anatomy-Aware Preset System}
\label{sec:presets}

We encoded 8 of the 75 atomic rhinoplasty sub-procedures from Daniel's
\emph{Mastering Rhinoplasty} taxonomy~\citep{daniel2010mastering} that are
detectable from a single frontal photograph: dorsal hump reduction, tip
narrowing, bridge narrowing, alar base narrowing, dorsal straightening,
tip definition, tip rotation up, and nose shortening. Each sub-procedure has a landmark-measurement detection threshold
and a prompt fragment. The prompt builder selects the top 3 detected
sub-procedures by a fixed clinical-priority ordering (PRIORITY\_ORDER in the released code and data, derived from Daniel's taxonomy frequency rankings) and concatenates their fragments,
producing a subject-specific prompt. The blepharoplasty (8) and
rhytidectomy (8) preset definitions follow the same architecture using
Tessier orbital and SMAS-layer anatomy respectively; they are released as part of the
framework. Here Daniel's taxonomy refers to the atomic sub-procedure
catalogue of \emph{Mastering Rhinoplasty}~\citep{daniel2010mastering},
Tessier refers to the standard periorbital classification of eyelid and
brow anatomy, and SMAS denotes the superficial musculoaponeurotic system.
We note that the SMAS itself is not visible in a frontal photograph; the
rhytidectomy presets therefore act on its surface correlates (jawline and
platysmal-band contour) rather than on the deep layer directly, and the
blepharoplasty presets model upper-eyelid changes only.

\subsection{Diffusion Inpainting}

FLUX.1-Fill-dev~\citep{flux2024}, a 12-billion-parameter rectified flow fill model, is
used with a pretrained depth ControlNet~\citep{zhang2023controlnet}
(the released code and data) at conditioning scale 0.5.
The \texttt{FluxInpaintPipeline} uses guidance scale 3.5, 20 denoising steps,
BF16 precision, and $512 \times 512$ resolution. Inpainting strength for
rhinoplasty is 0.75 at 50\% depth modification intensity. No procedure-specific
training is required for the base pipeline (a separate Direct Preference Optimization (DPO) LoRA probe is reported in Appendix~\ref{app:dpo_hparams}).

\subsection{Mask-Decomposed Evaluation: SurgicalScore}
\label{sec:eval_metric}

Standard full-face metrics conflate two distinct objectives: outside-mask
fidelity to the input, and inside-mask accuracy against the ground-truth
postoperative image. We define \textbf{SurgicalScore}, which decomposes
evaluation by mask region and reference:
\begin{itemize}
\item \textbf{Outside-mask SSIM}: prediction vs.\ input on pixels where
  $M < 0.5$. Approaches 1.0 for methods preserving non-surgical regions.
\item \textbf{Inside-mask LPIPS}: prediction vs.\ ground truth on the mask
  bounding crop (AlexNet backbone~\citep{zhang2018lpips}).
\item \textbf{Input ArcFace}: output vs.\ input cosine similarity (identity
  preservation).
\item \textbf{GT ArcFace}: output vs.\ ground truth cosine similarity (postoperative target-proximity proxy in identity space).
\item \textbf{BL ArcFace}: input vs.\ ground truth (dataset-level baseline,
  not a method score).
\end{itemize}

Because the composite $C = M \odot G + (1-M) \odot I$ differs from the input
only on the masked region, the operational identity preservation reduces to
the directly measurable outside-mask SSIM, which exceeds $0.999$ across all
evaluated cohorts (a property of the compositing rule, not of any backbone).
Empirically, the local Lipschitz constant of ArcFace under masked perturbations
is bounded by $L_{p95} \leq 1.2 \times 10^{-5}$ ($N{=}255$ samples,
Appendix~\ref{app:sensitivity}).

A \textbf{Mask-Area Modifier} corrects small-mask inflation: metrics are computed
on a resize-to-256 crop of the mask bounding box to normalize the effective
contribution of the surgical region. The affine rescale (slope 0.70, intercept 0.30) anchors the
passthrough floor at 0.30, so any SurgicalScore below 0.30 is scored below
returning the input under the protocol. Component weights are protocol hyperparameters; the floor is set by the affine rescale, not by the weights.

This decomposition prevents the confound observed in prior work~\citep{landmarkdiff},
where over 95\% of the identity score originated from composited pixels
(0.509 composited vs.\ 0.023 without).

\paragraph{SurgicalScore composite.}
For scored validation cases, the five components are aggregated as:
\begin{equation}
\mathrm{SS} = 0.30 + 0.70 \cdot \frac{R_O - R_I}{1 - R_I}, \quad
R_O = 0.40\,A + 0.30\,B + 0.15\,C + 0.10\,D + 0.05\,E
\label{eq:ssv5}
\end{equation}
where $R_I$ is $R_O$ evaluated with $O = I$ (the passthrough reference),
and a hard ArcFace identity gate disqualifies candidates when
$\cos(I, O) < 0.65$ (Input ArcFace, input-to-output). Component weights balance directional
and fidelity terms; the passthrough floor at $0.30$ is set by the affine rescale (slope $0.70$,
intercept $0.30$). Both are protocol hyperparameters fixed before any cross-method comparison.
SurgicalScore is the localized-fidelity protocol used for cross-method comparison; the
GT-vs-BL ArcFace gap is reported separately as an independent identity-space diagnostic.
The five components: $A$ = directional alignment (cosine of morphometry edit vectors,
as in~\citet{gal2022stylegan_nada}); $B$ = edit magnitude fit (asymmetric log-ratio,
over-edit penalized at $1.5\times$, under-edit at $1.0\times$); $C$ = masked LPIPS
fidelity on a 256$\times$256 mask-crop resize (Mask-Area Modifier applied); $D$ =
face realism: mean of SER-FIQ~\citep{terhorst2020serfiq} and CR-FIQA~\citep{boutros2023crfiqa} quality scores, with a three-signal proxy (embedding norm, Laplacian variance, color balance) as fallback when FIQA models are unavailable; $E$ = outside-mask
soft preservation ($1 - \mathrm{LPIPS}_\text{out} / \tau$, $\tau{=}0.10$).
\citet{gal2022stylegan_nada} introduced directional cosine loss in image editing;
$A$ adapts this to procedure morphometry space.
The rhinoplasty morphometry vector $\phi_\text{morph}^\text{rhino} \in \mathbb{R}^5$ contains:
Goode ratio~\citep{goode1984rhinoplasty} (lateral-view approximation computed from frontal photographs: tip projection / nasal length);
nasolabial angle (lateral-view approximation computed from frontal photographs, subnasale, landmark 2); alar width / intercanthal distance;
dorsal line RMS deviation / nasal length; nasal length / face height.
All components are scale-invariant ratios or normalized angles.

\paragraph{SS\textsubscript{raw} calibration anchor.}
SurgicalScore is the calibrated metric used for all cross-method comparison.
The uncalibrated raw composite $R_O = 0.40\,A + 0.30\,B + 0.15\,C + 0.10\,D + 0.05\,E$
(without the passthrough-anchored affine rescale) is reported alongside as a calibration
check; we refer to it as SS\textsubscript{raw}.
Under SS\textsubscript{raw}, an oracle control in which the system output is replaced
directly by the ground-truth postoperative image (no hard-mask composite) achieves
$0.919$ $[0.918, 0.920]$ on the $N{=}211$ cohort ($B{=}10{,}000$ bootstrap).
This shows that a perfect-predictor control occupies the high end of the
SS\textsubscript{raw} range, consistent with intended use of the $[0,\,1]$ scale.
Aggregate \method{} SS\textsubscript{raw} on $N{=}211$ is $0.563$ $[0.536, 0.590]$.
SS\textsubscript{raw} values for all ablation variants appear in
Table~\ref{tab:composite_ablation}.

\paragraph{Weight rationale and the GT-paste paradox.}
The component weights place $70\%$ of the raw score on edit geometry
($A$, direction, $0.40$; $B$, magnitude, $0.30$) because the surgical
intent is geometric: a goal visualization is correct insofar as it moves
the nasal landmarks in the operated direction and to the operated degree.
The remaining weight grades whether an edit is genuine rather than a
texture artifact: in-mask LPIPS to ground truth ($C$, $0.15$), realism
($D$, $0.10$), and outside-mask preservation ($E$, $0.05$, kept low because
the composite already guarantees it by construction). These weights are
protocol hyperparameters fixed before any cross-method comparison, not fit
to favor \method{}, and method rankings are largely stable under weight
perturbation (Dirichlet resampling, Appendix~\ref{app:sensitivity}).
We report one limit of the protocol openly. Among the GT-paste controls
(Table~\ref{tab:composite_ablation}), substituting the real postoperative
image for the model output \emph{without} compositing is the legitimate
perfect-predictor anchor and scores SS\textsubscript{raw} $0.919$. The same
real outcome \emph{feathered into the input mask} (the operational
compositing path), however, scores only calibrated SS $0.546$
(SS\textsubscript{raw} $0.569$), below \method{}'s calibrated SS $0.599$.
Two mechanical effects cause this. First, the realism term penalizes real
clinical photographs: FIQA scores the genuine postoperative capture at
$D{=}0.696$ versus $0.835$ for a clean diffusion output, because real
captures carry lighting, expression, and capture-condition artifacts that
the diffusion prior smooths away. Second, the feathered seam perturbs
MediaPipe landmark estimation, depressing the geometry terms $A$ and $B$
relative to the no-composite paste. SurgicalScore therefore partially rewards idealized,
artifact-free appearance, the same tendency toward results more symmetric
than surgery achieves that our clinical reviewer noted
(Sections~\ref{sec:results} and~\ref{sec:limitations}). We accordingly treat
SurgicalScore as a \emph{diagnostic protocol for localized edit fidelity,
not a clinical-validity ground truth}: the no-composite GT-paste is the
SS\textsubscript{raw} ceiling, and the feathered GT-paste is a diagnostic control that
quantifies the protocol's sensitivity to seams and to FIQA's bias against
real photographs. Clinical validity requires the blinded multi-rater study
deferred to future work.

\paragraph{Component-level discrimination.}
Each of the five SurgicalScore components flags failures the others miss.
Case Nose\_33 receives outside-mask preservation score ($E$) $= 0.000$ from a
compositing artifact in the mask boundary region despite high directional alignment
($A = 0.997$) and edit magnitude ($B = 0.872$).
Case Nose\_154 receives directional alignment ($A$) $= 0.004$ from a preset selection
mismatch despite intact realism ($D = 0.738$) and outside-mask preservation
($E = 0.540$).
Case Nose\_161 receives edit magnitude ($B$) $= 0.049$ from insufficient in-mask change
despite high directional alignment ($A = 1.000$) and realism ($D = 0.736$).
These three cases are not discriminable on full-face ArcFace alone: their
$\cos(\text{input}, \text{output})$ scores are 0.740, 0.858, and 0.692,
respectively, all above the identity gate threshold.

\subsection{Decision Model}
\label{sec:decision_model}

Seven hard gates must be passed for a candidate to be valid: (1) identity:
Input ArcFace $\geq 0.65$; (2) outside-SSIM: $\geq 0.95$; (3) landmark
drift: $\leq 15.0$ px; (4) dark-hole: mask HSV-V$<$20 area $\leq 0.5\%$;
(5) color consistency: mean hue shift $\leq 15.0^\circ$; (6) crease: blepharoplasty only;
(7) procedure fidelity: landmark-atlas severity bands. Passing candidates are
ranked by a weighted composite (0.25 identity, 0.15 outside-SSIM, 0.15
landmark drift, 0.15 color consistency, 0.15 cross-method agreement, 0.10 TPS agreement,
0.05 aesthetic). Cross-method agreement is computed only among \method{}'s own
candidate seeds and the deterministic TPS/preset proxies; no baseline outputs
or postoperative ground truth are used for candidate selection. Each
candidate receives a verdict: \textsc{pass} (all 7 hard gates clear),
\textsc{borderline} (all gates clear but the weighted composite falls
within $0.05$ of the rejection floor), or \textsc{fail} (any hard gate
trips). If no diffusion candidate passes, the deterministic TPS warp is
substituted as a non-generative geometric fallback.

\section{Experiments}
\label{sec:experiments}

\subsection{Dataset}

We use two cohorts. The first is the HDA Plastic Surgery
Database~\citep{rathgeb2020plastic}, a public corpus of identity-paired
pre/post-operative facial photographs, used for matched pairs across
procedures. The second is an external rhinoplasty cohort drawn from the
American Society of Plastic Surgeons (ASPS) public photo gallery and a
private clinical archive (PCA), used for the headline comparison. The HDA
Plastic Surgery Database contains 638
subjects. The matched HDA test splits contain 27 blepharoplasty, 21 rhinoplasty,
and 9 rhytidectomy pairs. The headline rhinoplasty comparison
(Section~\ref{sec:results}, Table~\ref{tab:strict_n211}) uses the
$N{=}211$ ASPS/PCA cohort described below. The matched 21-pair HDA
rhinoplasty split is used for the DPO pilot probe (Appendix~\ref{app:dpo_hparams})
and for prior-work comparability in Appendix~\ref{app:n21_detailed}.

We further assembled an external validation corpus from ASPS Photo
Gallery~\citep{asps2024gallery} and an additional private clinical archive
of paired pre/post photographs (\emph{PCA}), filtered to frontal views
(InsightFace yaw-symmetry $\geq 0.82$) and pre/post identity floor
($\geq 0.35$ ArcFace cosine), yielding 457 rhinoplasty pairs. All external
pairs are held out from training, hyperparameter selection, and preset design.

HDA is the only public corpus with identity-paired pre/post photographs across
the three procedures; CelebA-HQ~\citep{karras2018celebahq} and
FFHQ~\citep{karras2019ffhq} contain none, commercial 3D platforms are
proprietary, and SurFace1259~\citep{varghaei2025surface} was unavailable.
The identity-floor filter excludes pairs whose appearance change exceeds
plausible surgical drift; \texttt{rhytidectomy\_Facelift\_08} is removed
after manual review identified two different subjects.

\subsection{Baselines}

Three instruction-conditioned baselines were evaluated under two protocols: the $N{=}211$ ASPS+PCA cohort (used for the headline comparison in Table~\ref{tab:strict_n211}) and the matched $N{=}21$ HDA paired pool (used for prior-work comparability in Appendix~\ref{app:n21_detailed}): (1)~\textbf{InstructPix2Pix}~\citep{brooks2023instructpix2pix},
a full-image text-instruction editor applied with the same preset-derived
prompts; (2)~\textbf{ICEdit}~\citep{zhang2025icedit}, FLUX.1-Fill-dev with a
MoE-LoRA via diptych inference; (3)~\textbf{FLUX.1-Kontext-dev}~\citep{flux_kontext_2025},
a text-conditioned image-to-image variant run zero-shot followed by the same
hard-mask composite used in \method{}.

Three internal diagnostic baselines accompany the main comparators: Direct Copy
(Input ArcFace $=1.0$ upper bound), TPS Warp (geometric morphing only), and
FLUX Inpainting without ControlNet.

\subsection{Implementation}

Inference runs on a single NVIDIA L40S GPU (48~GB) in BF16. Depth estimation
uses Depth Anything V2 Small~\citep{yang2024depth}. Landmarks use MediaPipe FaceLandmarker
(478 points; successor of the 468-vertex face mesh~\citep{kartynnik2019mediapipe}). ArcFace uses InsightFace buffalo\_l
(R50). LPIPS uses AlexNet~\citep{zhang2018lpips}. Inference takes approximately
20 seconds per image. Training details for the DPO LoRA are given in
Appendix~\ref{app:dpo_hparams}.

\subsection{Statistical Inference}

We computed 95\% confidence intervals (CIs) on reported means wherever per-case
or seed-level replicates were available, via percentile bootstrap ($B{=}10{,}000$
resamples, seed~42). For Table~\ref{tab:surgical_score}
\method{} rows, matched-$N$ per-case GT Arc scores were not available locally;
CIs were bootstrapped over five seed-level means (denoted $\dagger$). ICEdit
CIs were bootstrapped over per-case scores ($N{=}21$). IP2P and Kontext
per-case data were not available; their CIs are not reported.

Bootstrapping over seed-level means captures inference variance (sensitivity to
random initialization) but not case-level variance (sensitivity to which 21
cases constitute the test set). The reported $95\%$ CIs are therefore not estimates
of case-sampling uncertainty: they reflect seed-to-seed reproducibility, not
the full uncertainty that would be estimated from a larger rhinoplasty population.

Within-protocol Input ArcFace ordering: \method{} exceeded ICEdit on rhinoplasty
($p{=}0.0361$, $N_{\text{ours}}{=}16$, $N_{\text{ice}}{=}21$, two-sample 10,000-permutation, two-sided;
the $N$ asymmetry reflects InsightFace detection failures on $5$ of $21$ \method{} cases on the matched HDA pool, while the headline $N{=}211$ cohort retains all $211$ outputs).
This test is included for protocol consistency; cross-method ordering on Input ArcFace is
not informative of inside-mask quality, since the metric is structurally confounded by
hard-mask compositing. For the $N{=}211$ cohort, paired sign-flip permutation tests against each
baseline are reported in Table~\ref{tab:paired_n205} ($p < 10^{-4}$ all three).
For the matched $N{=}21$ DPO probe, per-case zero-shot GT ArcFace was not stored
at evaluation time, so the probe is directional only.

\section{Results}
\label{sec:results}

\paragraph{Cohorts.} External corpus assembly: 457 raw rhinoplasty pairs from
ASPS Photo Gallery ($N{=}441$) and PCA ($N{=}16$). Two filtered cohorts derive
from this pool: (a) the $N{=}211$ cohort (ASPS $=202$, PCA $=9$)
requires both MediaPipe landmark detection and pre/post ArcFace $\geq 0.65$;
this is the headline cohort. (b) The gate-pass $N{=}438$ subset
(ASPS $=422$, PCA $=16$) requires only ArcFace $\geq 0.65$ on at least one
of five seeds; reported as the external gate-pass rate in
Section~\ref{sec:surgeon-verification} and the source-stratified pooled
GT-vs-BL replication in Appendix~\ref{app:external_sources}. (c) Matched $N{=}21$ HDA rhinoplasty
pool, used for the DPO probe and Appendix~\ref{app:n21_detailed}.

\subsection{Headline ArcFace and SurgicalScore on Rhinoplasty}

\paragraph{ArcFace headline.}
Table~\ref{tab:strict_n211} gives the headline result on the $N{=}211$
rhinoplasty cohort: ASPS public gallery ($N{=}202$) and a private clinical
archive (PCA, $N{=}9$), filtered to cases passing MediaPipe landmark
detection and pre/post same-person ArcFace at $0.65$ cosine. Across all
four methods evaluated, no method achieves positive mean GT ArcFace gain
over the unedited preoperative input. \method{} achieves output-to-GT
cosine $0.662$ versus input-to-GT $0.711$, gap $-0.048$ ($95\%$ CI
$[-0.055, -0.042]$, $B{=}10{,}000$); ICEdit, Kontext, and IP2P gaps are
$-0.139$, $-0.242$, $-0.294$ with non-overlapping CIs
(Table~\ref{tab:strict_n211}). On the four-way detected intersection
($N{=}205$), \method{} reduces the gap by $+0.090$ vs.\ ICEdit, $+0.189$
vs.\ Kontext, and $+0.245$ vs.\ IP2P, all paired sign-flip permutation
$p < 10^{-4}$ (Table~\ref{tab:paired_n205}). The fraction of cases where
output exceeds input proxy under GT ArcFace is $16.1\%$ for \method{}
versus $0.0$/$4.3$/$1.0\%$ for the baselines.

\paragraph{SurgicalScore.}
Mean SurgicalScore on \method{} on this cohort is $0.599$ $[0.579, 0.619]$
vs.\ ICEdit $0.502$ $[0.467, 0.535]$, InstructPix2Pix $0.337$
$[0.295, 0.380]$, and FLUX.1-Kontext-dev $0.229$ $[0.188, 0.271]$
(Appendix~\ref{app:ss_strict}, Table~\ref{tab:ss_strict_n211}). On the
four-way detected intersection ($N{=}188$), \method{} exceeds ICEdit by
$\Delta{=}+0.091$, InstructPix2Pix by $\Delta{=}+0.264$, and Kontext by
$\Delta{=}+0.368$, all paired sign-flip permutation $p < 10^{-4}$. \method{}
threshold-pass rates: $208/211$ ($98.6\%$) clear $0.35$, $145/211$
($68.7\%$) clear $0.50$.

\paragraph{Operational.}
The pipeline verdict is \textsc{pass} on $208/211$ and \textsc{borderline}
on $3/211$. The TPS fallback was not substituted on any $N{=}211$
case ($0/211$); fallback fired on $3/57$ ($5.3\%$) of the matched HDA
pool. The $N{=}457$ external corpus passes the identity gate at $95.8\%$
($438/457$).

Outside-mask SSIM exceeds $0.999$ for \method{} by architectural
construction; the same property holds for Kontext under the same composite
(Appendix~\ref{app:composite_full}), confirming the preservation follows
from the compositing rule, not from any one backbone. Mask-crop LPIPS
replicates the ranking: \method{} matches input-proxy LPIPS within
$\pm 0.012$, while baselines drift $0.12$--$0.20$ further from GT
(Appendix~\ref{app:mask_crop}). The methodological diagnosis transfers
across procedures (Appendix~\ref{app:cross_proc}).

\begin{table}[t]
\centering
\small
\caption{\textbf{Main result on the $N{=}211$ rhinoplasty cohort}
(ASPS public gallery $N{=}202$ + PCA private clinical archive $N{=}9$,
passing MediaPipe landmark detection and same-person ArcFace at $0.65$
cosine). \textbf{No method achieves positive mean GT ArcFace gain over the
unedited preoperative input.} \method{} achieves the smallest gap of any
method evaluated, with $16.1\%$ of cases positive ($0.0$/$4.3$/$1.0\%$ for
ICEdit/Kontext/IP2P). Per-method $N$ varies
due to InsightFace detection failures on a subset of generated outputs;
BL Arc is computed on the corresponding per-method input subset, so each
method's gap is paired within its own attrition mask. Gap 95\% CIs are
percentile bootstrap over per-case scores ($B{=}10{,}000$, seed~42).
\%~Pos.\ is the fraction of cases where output-to-GT ArcFace exceeds
input-to-GT ArcFace; \method{} is the only method exceeding $5\%$.
Paired statistical comparisons against \method{} are reported in
Table~\ref{tab:paired_n205}.}
\label{tab:strict_n211}
\begin{tabular}{@{}lccccc@{}}
\toprule
\textbf{Method} & \textbf{N} & \textbf{GT Arc}$\uparrow$ & \textbf{BL Arc} & \textbf{Gap [95\% CI]} & \textbf{\% Pos.}$\uparrow$ \\
\midrule
InstructPix2Pix     & 208 & 0.417 & 0.711 & $-0.294$~$[-0.331, -0.256]$ & 1.0 \\
ICEdit              & 206 & 0.573 & 0.712 & $-0.139$~$[-0.148, -0.130]$ & 0.0 \\
FLUX.1-Kontext-dev  & 211 & 0.469 & 0.711 & $-0.242$~$[-0.258, -0.225]$ & 4.3 \\
\method{} (ours)    & 211 & \best{0.662} & 0.711 & \best{$-0.048$~$[-0.055, -0.042]$} & \best{16.1} \\
\bottomrule
\end{tabular}
\end{table}

\begin{table}[t]
\centering
\small
\caption{\textbf{Paired comparison against \method{} on the four-way detected
intersection ($N{=}205$).} Each baseline is restricted to the cases where all
four methods produced a detected face, and the per-case ArcFace gap (output-to-GT
minus input-to-GT) is differenced row-wise against \method{}. Positive $\Delta$
favors \method{}. CI is percentile bootstrap over per-case differences
($B{=}10{,}000$). $p$ is a paired sign-flip permutation test
($B{=}10{,}000$). All three baselines are rejected against \method{} at
$p < 10^{-4}$.}
\label{tab:paired_n205}
\begin{tabular}{@{}lccc@{}}
\toprule
\textbf{Comparison (\method{} $-$ X)} & \textbf{$\Delta$ Gap} & \textbf{95\% CI} & \textbf{$p$ (paired perm.)} \\
\midrule
\method{} $-$ ICEdit            & $+0.090$ & $[+0.079, +0.101]$ & $<10^{-4}$ \\
\method{} $-$ FLUX.1-Kontext-dev & $+0.189$ & $[+0.172, +0.205]$ & $<10^{-4}$ \\
\method{} $-$ InstructPix2Pix    & $+0.245$ & $[+0.207, +0.285]$ & $<10^{-4}$ \\
\bottomrule
\end{tabular}
\end{table}

\paragraph{External cohort and surgeon review.}\label{sec:surgeon-verification}
On the $457$-pair external corpus, pooled GT Arc $0.597$ vs.\ BL Arc
$0.664$ (gap $-0.067$ vs.\ $-0.048$ on $N{=}211$, Appendix~\ref{app:external_sources}):
the negative gap holds on the broader cohort, larger magnitude consistent
with more difficult cases. A board-certified surgeon judged six outputs
plausible while flagging over-symmetrization; multi-rater blinded assessment with
the Global Aesthetic Improvement Scale (GAIS) and Rhinoplasty Outcome Evaluation
(ROE) is future work.

\subsection{The Composite Alone Does Not Explain the Lead}
\label{app:composite_ablation}

To isolate whether \method{}'s SurgicalScore lead comes from the hard-mask
composite step alone (versus the FLUX-Fill backbone choice and depth
conditioning), we applied the same composite operation post-hoc to ICEdit and
InstructPix2Pix outputs on the $N{=}211$ cohort:
$C = M \odot \text{baseline} + (1-M) \odot \text{input}$. We additionally
constructed GT-paste controls, $C = M \odot \text{GT} + (1-M) \odot \text{input}$,
to test how SurgicalScore responds to direct mask transfer of postoperative
pixels under the same compositing operation. These are diagnostic paste
controls, not upper bounds, because seams, lighting mismatch, expression
mismatch, and registration drift affect landmark and realism components.
All variants were scored with the identical SurgicalScore protocol used for
the headline numbers.

\begin{table}[h]
\centering
\small
\caption{Composite-equipped baseline ablation on $N{=}211$. Mean
SurgicalScore $[95\%\,$CI$]$ from per-case bootstrap ($B{=}10{,}000$);
SS\textsubscript{raw} is the uncalibrated raw composite $R_O$ (no passthrough-anchored
affine rescale); component breakdown shown. \textbf{Equipping IP2P and ICEdit
with the same feathered hard-mask composite as \method{} closes most of the
bare-baseline gap, yet \method{} retains a paired-significant lead}: \method{}
$-$ IP2P$+$composite $=+0.045$ $[+0.013, +0.077]$, $p{=}0.008$; \method{}
$-$ ICEdit$+$composite $=+0.058$ $[+0.026, +0.090]$, $p{<}10^{-3}$
(paired permutation $B{=}10{,}000$ on the per-method detected intersection).
The \emph{GT-paste control} rows are diagnostic
paste tests, not upper bounds: they show how the metric responds to direct
mask transfer of postoperative pixels under different composite operations,
and reveal that compositing seams (hard or feathered) drift MediaPipe
landmark predictions. With no compositing the GT paste maximizes
$A{=}B{=}C{=}1.000$ but loses outside preservation ($E{=}0.000$, SS $=0.703$,
SS\textsubscript{raw} $=0.919$); under a feathered GT-paste composite
($\sigma{=}12$ for this diagnostic control, distinct from the operational
$\sigma{=}15$ mask of Section~\ref{sec:methods}) the seam
recovers $E{=}0.989$ but A and B drop to $\approx0.36$--$0.52$
(SS $=0.546$, SS\textsubscript{raw} $=0.569$). \method{}'s $0.599$
exceeds the feathered-paste score because FLUX-Fill achieves higher realism
($D{=}0.835$ vs.\ $0.696$) and slightly larger landmark magnitude
($B{=}0.395$ vs.\ $0.361$). The lower $D{=}0.685$ on the GT-paste-no-composite
control reflects an apparent FIQA asymmetry on clinical photographs: real
postoperative captures carry lighting, expression, and capture-condition
artifacts that FIQA scores below clean diffusion outputs, so $D$ should be
read comparatively across rows rather than as an absolute ceiling. Source: released code and data.}
\label{tab:composite_ablation}
\resizebox{\columnwidth}{!}{%
\begin{tabular}{@{}lcccccccc@{}}
\toprule
\textbf{Variant} & \textbf{N} & \textbf{Mean SS [95\% CI]} & \textbf{SS\textsubscript{raw}} & \textbf{A} & \textbf{B} & \textbf{C} & \textbf{D} & \textbf{E} \\
\midrule
GT-paste control (no composite) & 211 & $0.703$ $[0.649, 0.756]$ & $0.919$ & 1.000 & 1.000 & 1.000 & 0.685 & 0.000 \\
\method{} (full)                & 211 & \best{$0.599$ $[0.579, 0.619]$} & $0.563$ & 0.524 & 0.395 & 0.662 & 0.835 & 1.000 \\
IP2P + feathered composite      & 209 & $0.565$ $[0.538, 0.591]$ & $0.521$ & 0.514 & 0.376 & 0.559 & 0.695 & 0.990 \\
ICEdit + feathered composite    & 208 & $0.549$ $[0.524, 0.575]$ & $0.501$ & 0.495 & 0.337 & 0.553 & 0.695 & 0.993 \\
GT-paste control (feathered)    & 207 & $0.546$ $[0.510, 0.582]$ & $0.569$ & 0.524 & 0.361 & 0.880 & 0.696 & 0.989 \\
GT-paste control (hard-mask)    & 207 & $0.543$ $[0.507, 0.578]$ & $0.567$ & 0.516 & 0.366 & 0.880 & 0.696 & 0.990 \\
ICEdit (no composite)           & 199 & $0.502$ $[0.467, 0.535]$ & $0.502$ & 0.568 & 0.448 & 0.478 & 0.691 & 0.000 \\
IP2P (no composite)             & 197 & $0.337$ $[0.295, 0.380]$ & $0.457$ & 0.531 & 0.337 & 0.487 & 0.708 & 0.000 \\
\bottomrule
\end{tabular}%
}
\end{table}

\paragraph{Discussion and limitations.}\label{sec:discussion}\label{sec:conclusion}
\method{} is a pre-consult goal-visualization tool, not a patient-specific
outcome predictor; the all-negative ArcFace gap across methods shows
that full-face identity verification is poorly aligned with localized
surgical edits under hard compositing (we demonstrate this for rhinoplasty;
we do not claim it for all edit types).
Open problem: a deployable non-oracle ranker (Appendix~\ref{app:deployable_ranker};
full limitations Section~\ref{app:limitations_full}).

\section{Limitations}
\label{app:limitations_full}
\label{sec:limitations}

\textit{Statistics.} $N{=}211$ uses per-case bootstrap and paired
permutation; matched $N{=}21$ DPO probe is directional only because
zero-shot per-case GT ArcFace was not stored.

\textit{Clinical evaluation depth.} Aesthetic plausibility was assessed by
a single board-certified facial plastic surgeon coauthor on six
representative outputs. This is a feasibility check, not a clinical study.
A multi-rater blinded panel evaluation using validated instruments (GAIS,
ROE) on a larger output sample is committed pre-deployment work.

\textit{2D representation constraints.} The pipeline operates on a single
frontal photograph and a 2D monocular depth map. This representation
cannot disambiguate bony vs.\ cartilaginous nasal deformation, model
surgeon-specific structural choices (e.g., spreader-graft placement,
septal reconstruction), or render lateral or oblique views. The qualitative failure mode of rhinoplasty volumetric miss reflects
this constraint: cases requiring full 3D shape change exceed what 2D
depth conditioning can encode.

\textit{Consent and governance.} HDA is used under its biometric-research
license; ASPS public-gallery and PCA clinical-archive images are used for
held-out evaluation only without redistribution. Released artifacts are
code, evaluation harnesses, preset configurations, and split manifests.
Clinical deployment beyond goal-visualization scope requires multi-site
IRB review.

\textit{Model-license compliance.} FLUX.1-dev and FLUX.1-Fill-dev are
distributed under the FLUX.1 [dev] non-commercial license, whose
restricted-use list includes biometric processing of FLUX-derived data.
A strict reading of this clause covers any biometric scoring of FLUX
outputs; we acknowledge that reading. Our use of these weights is
research-only and non-commercial: the pipeline is run inside a
controlled evaluation harness on retrospective de-identified or
public-release facial photographs, no FLUX-derived weights are
redistributed, and ArcFace and MediaPipe operate solely as
evaluation-side identity gates and landmark extractors during research
scoring rather than as components of any identification or enrollment
system. Generated outputs are simulation-only visualizations and are
not used for verification, enrollment, or identity-decisioning. Any
production deployment that involves identity verification on
FLUX-generated outputs would require a separate license assessment.

\textit{Anchoring risk.} Photorealistic outcome visualizations can anchor
patient expectations regardless of framing; the symmetry bias the surgeon
identified is one manifestation. Clinical deployment must pair the
visualization with explicit verbal anchoring during counseling.

\textit{Ranker design and appearance norms.} The K$=5$ oracle result
implies that future work will design a non-oracle candidate ranker. A
ranker trained on aesthetic preferences, surgeon ratings, or any reward
signal that encodes appearance norms would risk filtering candidates
toward an idealized phenotype rather than the patient-specific surgical
outcome. Ranker design must explicitly address this risk: audit ranker
preferences for race, gender, age bias; require multiple suggestions
surfaced rather than a single recommended one.

\textit{Fairness power analysis.} The Monk Skin Tone stratification
reports $N{=}4$ on tone~$7$, which is too small for a powered claim.
Under $\alpha{=}0.05$ and a two-sided test for a SurgicalScore difference
of $0.05$ against the N=211 variance, characterizing the gap with
$80\%$ power requires $N{=}25$--$40$ per stratum on tones $7$--$10$;
recruiting that cohort is committed pre-deployment work.

\bibliographystyle{plainnat}
\bibliography{refs}

\appendix
\section{Deployable K$=5$ Ranker Probe}
\label{app:deployable_ranker}

The K$=5$ best-of-$5$ oracle assumes GT access at scoring time. To assess
whether a deployable (no-GT) ranker can recover the oracle gain, we
evaluated six GT-free per-seed signals as candidate rankers on the
five-seed-complete subset ($N{=}207$): random selection, single fixed seed
(K$=1$ headline), max ArcFace$(out, in)$ (identity preservation), max
realism component~D, ArcFace$\times$D, and max outside-mask SSIM.
None recovered any of the K$=1{\rightarrow}$K$=5$ oracle gain
($+0.134$); all underperformed K$=1$ by $0.01$--$0.04$. Detail in
Table~\ref{tab:deployable_ranker}.

\begin{table}[h]
\centering
\small
\caption{Naive no-GT rankers vs.\ K$=5$ oracle on $N{=}207$ (Envisage
five-seed-complete cases). Per-case oracle selection by true SurgicalScore
raises the mean from $0.609$ (K$=1$) to $0.743$ ($+0.134$). Deployable
rankers based on identity preservation ($\arccos(O,I)$), realism (D),
outside-mask SSIM (E), or combinations all score below the single-seed
headline; the oracle gain comes from GT-aware components (A directional,
B magnitude, C masked LPIPS) that no-GT signals do not approximate.
A learned non-oracle ranker trained on (image, predicted SurgicalScore)
pairs is the natural next step; this is future work.}
\label{tab:deployable_ranker}
\begin{tabular}{@{}lcc@{}}
\toprule
\textbf{Ranker} & \textbf{Mean SS} & \textbf{$\Delta$ vs.\ K$=1$} \\
\midrule
Random seed                         & $0.574$ & $-0.035$ \\
\textbf{K$=1$ (seed $42$, headline)} & $\mathbf{0.609}$ & $0.000$ \\
$\arccos(O,I)$ (identity preservation) & $0.597$ & $-0.012$ \\
D (realism)                         & $0.580$ & $-0.029$ \\
$\arccos(O,I) \cdot D$              & $0.600$ & $-0.009$ \\
E (outside-mask SSIM)               & $0.595$ & $-0.014$ \\
\midrule
K$=5$ oracle (max true SS)          & $\mathbf{0.743}$ & $\mathbf{+0.134}$ \\
\bottomrule
\end{tabular}
\end{table}

\paragraph{Per-ranker interpretation.}
Each ranker's failure mode reflects a specific GT-blind tradeoff.
$\arccos(O,I)$ maximizes input--output similarity, which selects for
\emph{minimal} edit, the opposite of what the metric rewards (directional
alignment $A$ and magnitude $B$). $D$ (realism via FIQA) rewards
photorealistic outputs but does not correlate with surgical correctness; a
high-realism candidate may edit in the wrong direction.
$\arccos(O,I) \cdot D$ combines two GT-blind signals without resolving the
underlying GT-blindness. $E$ (outside-mask SSIM) is nearly identical
across seeds because the composite preserves outside-mask pixels by
construction, so the signal carries almost no per-seed information. Random
selection underperforms by construction. The common failure mode: GT-aware
components ($A$ directional, $B$ magnitude, $C$ masked LPIPS) drive the
$+0.134$ oracle gain, and no GT-free signal approximates them.

\paragraph{Learned-ranker sketch.}
A learned non-oracle ranker
$f(I_{\text{candidate}}, I_{\text{input}}, \ell_{\text{edit}}) \to \widehat{SS}$
predicting SurgicalScore from image and landmark features could
potentially recover the oracle gain at deployment. Training data already
exists: the $5$-seed $\times$ $N{=}211$ sweep yields $1{,}055$
(candidate, SurgicalScore) pairs, sufficient for fitting a small
regression head on pretrained features (e.g., DINOv2~\citep{oquab2024dinov2}
$+$ MLP, or a fine-tuned ViT). Held-out $N{=}21$ HDA matched cases provide
independent validation. The architectural separation between substrate
(this paper) and ranker (future work) means a successful ranker is a pure
post-hoc improvement: it does not require retraining the diffusion
backbone, and the current pipeline's K$=1$ outputs serve as the floor any
deployable ranker must clear. Open question: whether the (image, predicted
$SS$) supervision signal is dense enough to learn the GT-aware components
implicit in $A$, $B$, and $C$, or whether ranker training requires
additional clinical labeling.

\section{ArcFace as Identity Gate, Not Surgical-Fidelity Metric}
\label{app:arcface_gate}

ArcFace is trained for full-face verification and compresses the entire
face into a $512$-d embedding dominated by stable features (eye geometry,
jawline curvature, skull shape, hairline). The embedding mixes a small
localized change with the much larger global identity signal, so changes
at the scale of surgical rhinoplasty (mean mask area $3.65\%$ of pixels
across the $N{=}211$ cohort, range $1.32$--$10.90\%$) produce signal that
the metric registers but cannot attribute to surgical correctness. The same postoperative target is identified
as the same person as the preoperative input at high cosine ($0.71$)
precisely because rhinoplasty preserves global identity. The $-0.048$
N=211 gap reflects two things: a small ArcFace noise floor under
any pixel-level perturbation of the face (consistent across all four
tested methods, see paired comparisons in Table~\ref{tab:paired_n205}),
and a small diffusion artifact in the inpainted region that the metric
notices without being able to attribute to whether the change is
surgically correct. Per-case oracle selection over five seeds reduces the
gap by $73\%$ to $-0.015$ and produces positive identity gain on $33.9\%$
of cases (Appendix~\ref{app:best_of_k}), which is the correct read of the
candidate-space ceiling under our pipeline. We retain ArcFace as a strict
identity gate (output must remain identifiable as the same person,
$\cos\!\geq\!0.65$), which it is well-suited for, but not as a primary
outcome metric, which it was never designed to be. SurgicalScore,
landmark-derived edit vectors, and masked LPIPS serve the
continuous outcome-grading role; the two roles are complementary, not
redundant.

\clearpage
\section{SurgicalScore Decomposition Figure}
\label{app:decomposed_fig}

\begin{figure}[H]
  \centering
  \includegraphics[width=0.85\linewidth]{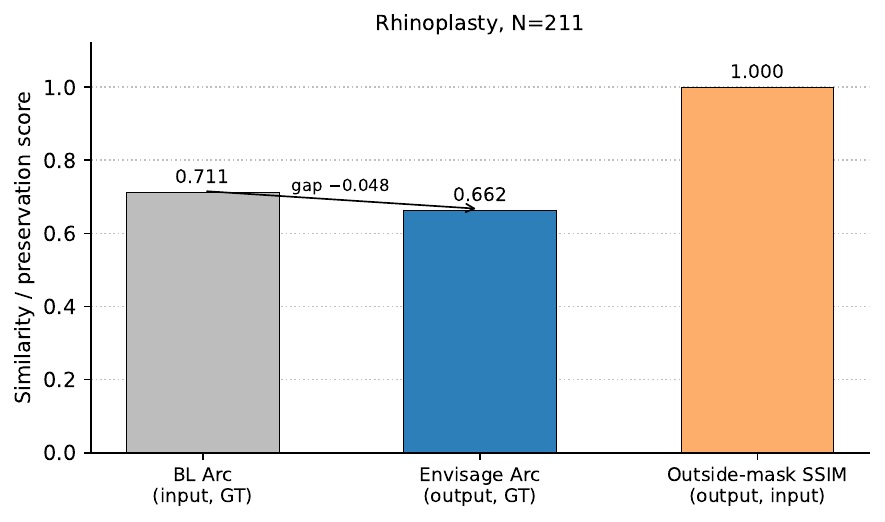}
  \caption{\textbf{Diagnostic comparison of full-face identity similarity and
    non-surgical-region preservation on the $N{=}211$ rhinoplasty
    cohort.}
    The non-surgical region is preserved by the hard-mask composite, so high
    full-face identity scores are dominated by copied pixels and should not be
    interpreted as surgical accuracy. This is a preservation diagnostic, not an
    additive decomposition of ArcFace (full-face ArcFace cosine is not
    region-additive). \method{} achieves output-to-GT ArcFace $0.662$ versus
    the input-to-GT proxy $0.711$ (gap $-0.048$); outside-mask SSIM is $1.000$
    by construction. The GT-below-BL gap is the central open problem, and
    SurgicalScore (Section~\ref{sec:eval_metric}) provides the mask-decomposed
    metric we use for cross-method comparison. Cross-procedure validation
    appears in Appendix~\ref{app:cross_proc}.}
  \label{fig:decomposed_arcface}
\end{figure}

\clearpage
\section{Qualitative Rhinoplasty Outputs}
\label{app:qualitative_fig}

\begin{figure}[H]
  \centering
  \includegraphics[width=0.65\linewidth,keepaspectratio]{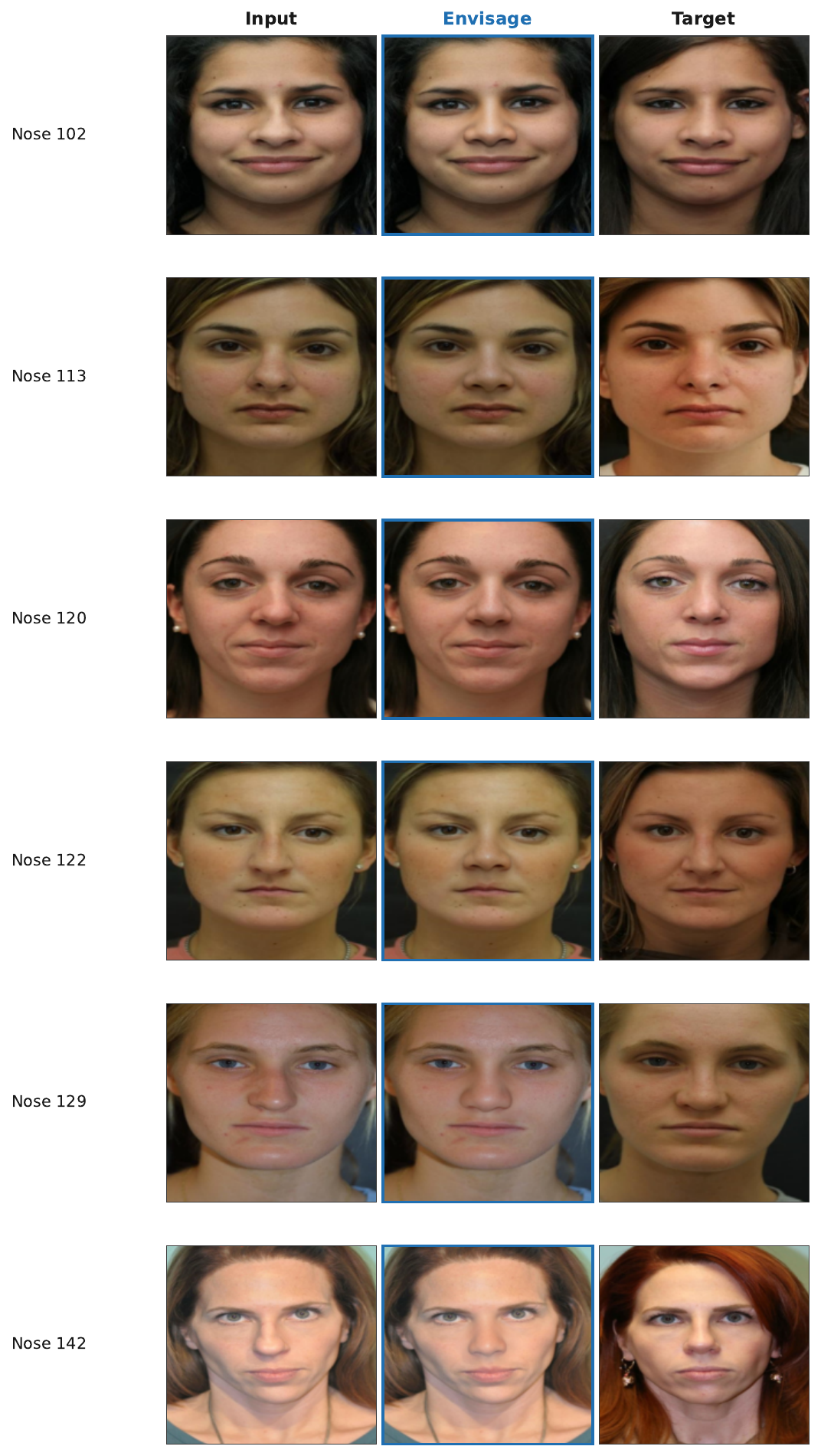}
  \caption{\textbf{Six surgeon-reviewed rhinoplasty cases} (Nose\_102,
    113, 120, 122, 129, 142). Columns: input, \method{} prediction,
    postoperative target. The board-certified facial plastic surgeon
    coauthor judged each as a goal-setting visualization on the matched
    HDA pool. Inside the mask the pipeline produces anatomically
    plausible edits; outside the mask identity preservation is exact by
    construction. Nose\_120 illustrates the symmetry-bias caveat
    described in Section~\ref{app:limitations_full}.}
  \label{fig:qualitative_rhinos}
\end{figure}

\begin{figure}[H]
  \centering
  \includegraphics[width=0.85\linewidth,keepaspectratio]{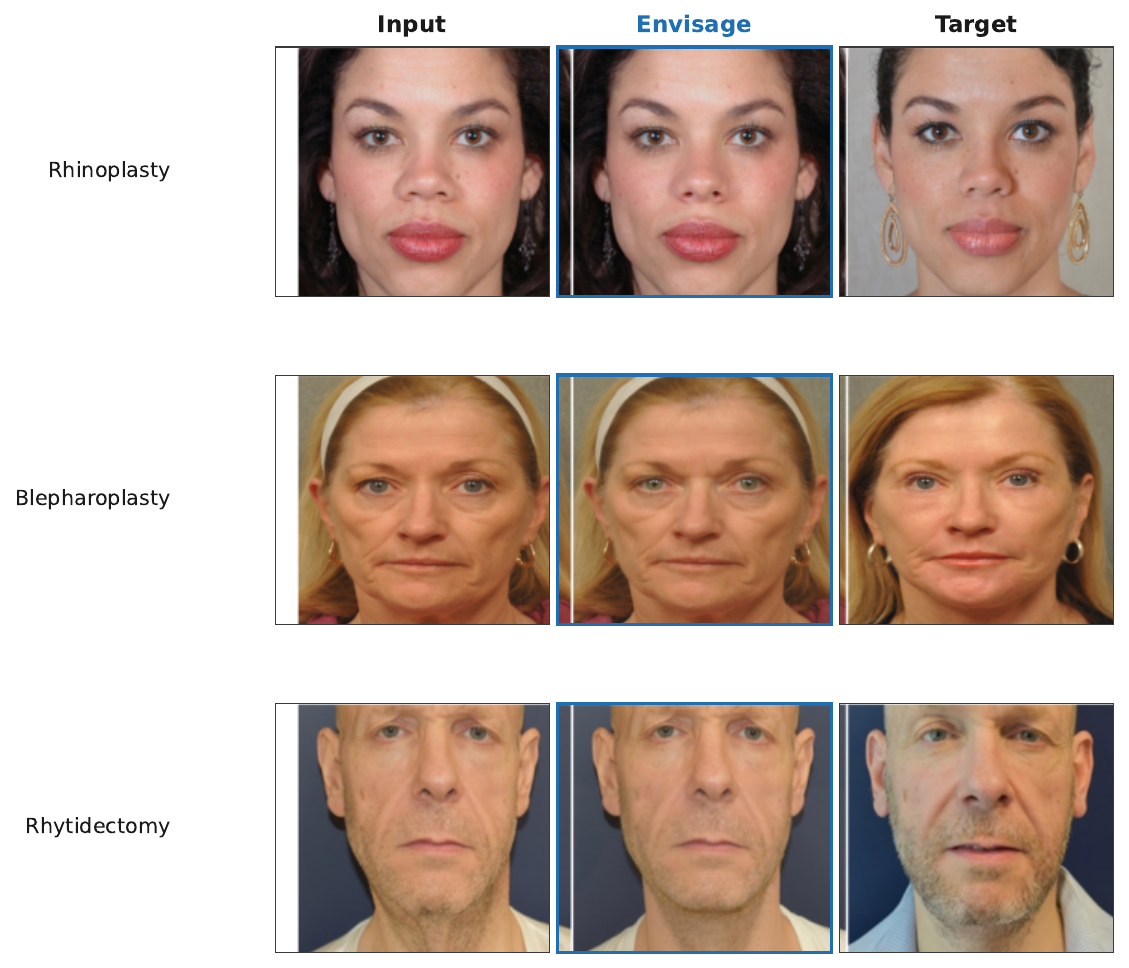}
  \caption{\textbf{Cross-procedure qualitative examples} on three
    subjects, using the released preset framework. Rows show
    rhinoplasty, blepharoplasty, and rhytidectomy. Columns show input,
    \method{} output, and target. All cells are rendered from the
    identical frontal pose, so differences reflect preset-level edits
    and subject-level anatomy. These cross-procedure examples are
    qualitative framework demonstrations; headline evaluation is
    rhinoplasty-specific.}
  \label{fig:other_procedures}
\end{figure}

\section{Detailed N=21 Matched HDA Paired-Pool Metrics}
\label{app:n21_detailed}

On the matched $N{=}21$ HDA rhinoplasty paired pool, \method{} produces the
smallest GT-vs-BL ArcFace gap of any method evaluated ($-0.077$ vs.\ ICEdit
$-0.167$, IP2P $-0.084$, Kontext $-0.245$), replicating the headline ranking
on a smaller paired cohort. SurgicalScore on this pool: \method{} $0.636$,
FLUX-Kontext-dev $0.600$, ICEdit $0.565$, InstructPix2Pix $0.545$ (the
matched $N{=}21$ Envisage composite of $0.636$ is footnoted on the
rhinoplasty Envisage row of Table~\ref{tab:cross_proc_baselines}, which
otherwise reports the headline $N{=}211$ cohort for cross-procedure
comparison).
Headline results in the main paper use the larger $N{=}211$ cohort
(Table~\ref{tab:strict_n211}) where all four methods produce outputs on the
same filtered case set.

\begin{table}[h]
\centering
\small
\caption{Detailed metrics on the matched $N{=}21$ HDA paired pool. All four
methods use the same preset-derived prompts. ArcFace metrics use InsightFace
detection subsets; shared-pool denominator reported. BL Arc: input vs.\ ground
truth (dataset property, not a method score; 0.687 for this pool). Out SSIM$^\ast$:
post-composite for \method{}. The Kontext row reports the value from the original baseline run, where the post-composite was applied with a different feathering radius; under \method{}'s exact compositing pipeline Kontext also reaches $>0.999$ (Appendix~\ref{app:composite_full}). IP2P and ICEdit do not perform mask compositing in their original formulations; their raw backbone-output SSIM is shown for reference and is not directly comparable to the composited rows. \method{}
95\% CIs: bootstrapped over 5 seed-level means ($\dagger$); this captures
inference variance only, not case-level variance. Traced to
the released code and data.}
\label{tab:surgical_score}
\resizebox{\columnwidth}{!}{%
\begin{tabular}{@{}lcccccc@{}}
\toprule
\textbf{Method} & \textbf{N} & \textbf{Out SSIM}$\uparrow$ &
\textbf{Input Arc}$\uparrow$ & \textbf{GT Arc}$\uparrow$ &
\textbf{BL Arc} & \textbf{In LPIPS}$\downarrow$ \\
\midrule
IP2P            & 21 & 0.835 & \best{0.914} & 0.603 & 0.687 & -- \\
ICEdit          & 21 & 0.710 [0.691, 0.730] & 0.765 [0.744, 0.785] & 0.520 [0.451, 0.571] & 0.687 & -- \\
Kontext         & 21 & 0.996$^\ast$ & 0.644 & 0.442 & 0.687 & 0.229 \\
\method{} (ours) & 21 & 1.000 [1.000, 1.000]$^\ast$$\dagger$ & 0.848 [0.841, 0.856]$\dagger$ & \best{0.610 [0.608, 0.613]$\dagger$} & 0.687 & 0.153 [0.149, 0.157]$\dagger$ \\
\bottomrule
\end{tabular}%
}
\end{table}

\section{Blepharoplasty and Rhytidectomy Preset Definitions}
\label{app:other_presets}

The \method{} framework includes 8 blepharoplasty and 8 rhytidectomy
sub-procedure presets following the same architecture as the rhinoplasty
presets (Section~\ref{sec:presets}). Each preset specifies (a) a landmark
detection threshold derived from procedure-specific morphometry; (b) a
depth modification Gaussian centered at landmark indices; (c) a TPS
displacement field; and (d) a prompt fragment composed with a base opening
and closing.

\paragraph{Blepharoplasty presets} (Tessier orbital taxonomy):
(1)~\texttt{upper\_skin\_excision} (upper-lid skin excess reduction),
(2)~\texttt{crease\_restoration} (supratarsal crease re-formation),
(3)~\texttt{upper\_dehooding} (severe-mode hood release on dermatochalasis),
(4)~\texttt{lid\_symmetry} (left/right palpebral fissure equalization),
(5)~\texttt{fat\_pad\_reduction} (medial and central fat pad reduction),
(6)~\texttt{lower\_bag\_reduction} (lower-lid bag flattening),
(7)~\texttt{tear\_trough\_smoothing} (tear-trough deformity correction),
(8)~\texttt{crow\_feet\_softening} (lateral periorbital rhytid softening).

\paragraph{Rhytidectomy presets} (SMAS-layer anatomy):
(1)~\texttt{jawline\_straightening} (mandibular contour redefinition),
(2)~\texttt{jowl\_elimination} (mandibular jowl resection at gonial angle),
(3)~\texttt{neck\_smoothing} (anterior cervical skin tightening),
(4)~\texttt{marionette\_softening} (marionette-line softening at oral
commissure), (5)~\texttt{platysmal\_band\_removal} (anterior platysmal
band corset plication), (6)~\texttt{prejowl\_correction} (pre-jowl sulcus
volumization), (7)~\texttt{submental\_definition} (submental angle
sharpening), (8)~\texttt{nasolabial\_softening} (nasolabial fold
softening).

\paragraph{Worked example: \texttt{upper\_skin\_excision}.}
Landmark indices: MediaPipe upper-lid landmarks $159$, $145$, $173$, $157$,
$158$, $160$, $161$, $246$ (right eye) plus mirror set (left eye); detection
threshold: upper-lid skin redundancy ratio $\geq 0.6$ (measured as ratio
of skin area above the supratarsal fold to lid-margin area). Depth
modification: 2D Gaussian kernel of $\sigma_x{=}18$\,px, $\sigma_y{=}9$\,px
centered at the supratarsal landmark, displacing depth by $-3$\,px (skin
removal). TPS displacement: upper-lid landmarks pulled superiorly by
$2$--$4$\,px scaled to measured redundancy. Prompt fragment: ``younger
upper eyelids with reduced skin excess, defined supratarsal crease.''
Full configurations including all four components per preset are in
the released code and data.

\section{DPO Training Hyperparameters and Pilot Probe Results}
\label{app:dpo_hparams}

DPO LoRA configuration: FLUX.1-Fill-dev, rank 16, $\beta{=}0.1$, learning rate $10^{-5}$,
2000 steps, cosine decay, BF16, single NVIDIA L40S. Preference pairs: winner/loser
inpaintings from HDA rhinoplasty training split, ranked by composite score
(identity 0.25, outside-SSIM 0.20, landmark drift 0.15). Reference model:
frozen SFT checkpoint. Implementation: the released code and data.

\begin{table}[h]
\centering
\small
\caption{Zero-shot \method{} vs.\ DPO-finetuned \method{} on $N{=}21$ rhinoplasty
(matched 5-seed protocol, identical hard-mask composite, identical 21-case set).
$\Delta$ GT$-$BL: identity-space target-proximity gap (negative = output further from GT than input).
Six of $21$ rhinoplasty cases flip to a positive delta on at least one DPO seed.
Pooled mean shows a directional shift on this small matched pool ($0.620$
DPO vs.\ $0.610$ zero-shot), but the GT-vs-BL gap remains negative for
both configurations ($-0.067$ DPO vs.\ $-0.077$ zero-shot); the
$N{=}211$ DPO sweep (Appendix~\ref{app:dpo_sweep}) finds no improvement
across six $\beta$ configurations, so this matched-pool result should
not be read as evidence that DPO improves \method{} at scale. Per-case
GT ArcFace was not stored for the zero-shot baseline, so no paired
permutation test is possible.
Source: released code and data.}
\label{tab:m9_dpo}
\begin{tabular}{@{}lccccc@{}}
\toprule
\textbf{Config} & \textbf{N} & \textbf{GT Arc}$\uparrow$ &
\textbf{BL Arc} & \textbf{$\Delta$ GT$-$BL} & \textbf{In LPIPS} \\
\midrule
Zero-shot (5-seed)        & 21 & $0.610 \pm 0.003$ & 0.687 & $-0.077$ & $0.153 \pm 0.004$ \\
DPO-finetuned (5-seed)    & 21 & $0.620 \pm 0.008$ & 0.687 & $-0.067$ & $0.158 \pm 0.001$ \\
\bottomrule
\end{tabular}
\end{table}

\section{External Validation Source Breakdown}
\label{app:external_sources}

The 457-pair external corpus combines two sources: the ASPS Photo Gallery
($N{=}441$ attempted), a public bank of patient outcome photographs
submitted by member surgeons, and a private clinical archive (PCA,
$N{=}16$ attempted) of paired pre/post photographs from a partnering
aesthetic practice. The two sources differ in curation: ASPS images are
surgeon-submitted with public-gallery framing (typically frontal,
well-lit, posed), while PCA images are clinical-record photographs with
consistent intra-patient capture conditions. PCA achieves a perfect
$16/16$ gate-pass rate; ASPS gate-pass is $422/441$ ($95.7\%$), with the
$19$ failures concentrated in cases where image-capture conditions
exceeded the InsightFace yaw-symmetry tolerance.

\begin{table}[h]
\centering
\small
\caption{External rhinoplasty validation by source. N attempted: total cases
submitted to the pipeline. N gate-pass: cases passing the 0.65 ArcFace
identity gate on at least one of 5 seeds. GT Arc and BL Arc are pooled
means over gate-pass cases. Traced to
the released code and data.}
\begin{tabular}{@{}lcccccc@{}}
\toprule
\textbf{Source} & \textbf{N attempted} & \textbf{N gate-pass} &
\textbf{Out SSIM} & \textbf{GT Arc} & \textbf{BL Arc} & \textbf{In LPIPS} \\
\midrule
ASPS  & 441 & 422 & 0.9998 & 0.595 & 0.660 & 0.240 \\
PCA   &  16 &  16 & 0.9998 & 0.657 & 0.712 & 0.231 \\
\midrule
Total & 457 & 438 & 0.9998 & 0.597 & 0.664 & 0.239 \\
\bottomrule
\end{tabular}
\end{table}

\section{Composite Ablation: Full Procedure Table}
\label{app:composite_full}

Table~\ref{tab:composite_full} reports outside-mask SSIM with and
without the hard-mask composite across all three procedures,
demonstrating that the $>0.999$ preservation property follows from the
compositing rule rather than backbone choice. Detailed cross-procedure
SurgicalScore breakdown is in Appendix~\ref{app:cross_proc}.

\begin{table}[h]
\centering
\small
\caption{Outside-mask SSIM with and without hard-mask composite across all three
procedures. Blepharoplasty and rhytidectomy results are included here for
framework completeness; they are not the primary evaluation target of this paper.
The Kontext backbone, when wrapped in \method{}'s exact compositing pipeline,
also produces outside-mask SSIM $>0.999$ on rhinoplasty (matched $N{=}21$ pool;
reported in Table~\ref{tab:surgical_score} body); the property follows from the
compositing rule and is not specific to one backbone.
Source: released code and data.}
\label{tab:composite_full}
\begin{tabular}{lcccc}
\toprule
& \multicolumn{2}{c}{\textbf{Composite ON}} & \multicolumn{2}{c}{\textbf{Composite OFF}} \\
\cmidrule(lr){2-3}\cmidrule(lr){4-5}
\textbf{Procedure} & \textbf{Envisage} & \textbf{$\sigma$} & \textbf{ICEdit} & \textbf{IP2P} \\
\midrule
Blepharoplasty & 0.9996 & 0.0000 & 0.650 & 0.681 \\
Rhinoplasty    & 0.9997 & 0.0000 & 0.710 & 0.835 \\
Rhytidectomy   & 0.9993 & 0.0001 & 0.651 & 0.837 \\
\bottomrule
\end{tabular}
\end{table}

\section{Cross-Procedure SurgicalScore Validation}
\label{app:cross_proc}

To assess whether SurgicalScore generalizes beyond rhinoplasty, we applied the
SurgicalScore protocol to the blepharoplasty and rhytidectomy subsets of HDA (full test split).
No re-training was performed; the same weights, landmarks, and calibration constants
were used. Components A--E were evaluated identically to the rhinoplasty pipeline;
only the procedure-specific morphometry vector $\phi_\text{morph}$ was swapped
to the corresponding blepharoplasty or rhytidectomy preset definition.

\begin{table}[h]
\centering
\small
\caption{SurgicalScore cross-procedure validation. Blepharoplasty and rhytidectomy
scores are computed on the full HDA test split ($N_\text{bleph}{=}51$, $N_\text{rhytid}{=}19$);
rhinoplasty uses the matched paired pool ($N{=}21$), where pre/post images are
both available for paired comparisons cited elsewhere in the paper. The
$27/21/9$ matched paired-pool splits are paired pre/post subsets of the full
test splits ($N_\text{bleph}{=}51$ vs.\ matched $N{=}27$, $N_\text{rhino}$ full
$=34$ vs.\ matched $N{=}21$, $N_\text{rhytid}{=}19$ vs.\ matched $N{=}9$).
Gate pass: ArcFace identity gate $\cos(I,O)\geq 0.65$.
Source: released code and data.}
\label{tab:cross_proc}
\begin{tabular}{@{}lcccccc@{}}
\toprule
\textbf{Procedure} & \textbf{N} & \textbf{Gate Pass} &
\textbf{Mean SS} & \textbf{SD} & \textbf{Max SS} &
\textbf{$N{\geq}0.35$} \\
\midrule
Blepharoplasty & 51 & 51/51 & 0.555 & 0.139 & 0.831 & 46/51 \\
Rhinoplasty    & 21 & 20/21 & 0.636 & 0.229 & 0.893 & 19/21 \\
Rhytidectomy   & 19 & 17/19 & 0.486 & 0.203 & 0.874 & 16/19 \\
\bottomrule
\end{tabular}
\end{table}

Blepharoplasty achieves 100\% identity gate pass rate, consistent with eyelid-only
masks that leave facial geometry intact. Rhytidectomy gate failures (2/19) correspond
to high-strength composite-lift cases where aggressive SMAS repositioning alters
facial geometry enough to drop ArcFace below the 0.65 threshold. Reported
mean SurgicalScore is $0.636$ for rhinoplasty (matched $N{=}21$ pool),
$0.555$ for blepharoplasty (full $N{=}51$), and $0.486$ for rhytidectomy
(full $N{=}19$); these denominators differ across procedures, so the
ranking is descriptive context rather than a like-for-like cross-procedure
comparison.

\begin{table}[h]
\centering
\small
\caption{Cross-procedure SurgicalScore. Baselines (ICEdit, IP2P,
FLUX.1-Kontext-dev) are scored on the matched HDA paired pool
($N_\text{bleph}{=}27$, $N_\text{rhino}{=}21$, $N_\text{rhytid}{=}9$).
\method{} is scored on the headline $N{=}211$ ASPS+PCA cohort for
rhinoplasty (where composite and per-component breakdown are jointly
computed), and on the full HDA test split for blepharoplasty ($N{=}51$)
and rhytidectomy ($N{=}19$); none of the Envisage rows in this table is
therefore directly paired with the baseline rows (different denominators)
and they should be read as descriptive cross-procedure context rather
than head-to-head comparison. The matched $N{=}21$ Envisage rhinoplasty
composite of $0.636$ is reported in Appendix~\ref{app:n21_detailed}.
Component E (outside-mask
preservation) is the diagnostic that consistently separates \method{}
($\approx 1.0$ on rhinoplasty, $\approx 0.55$ on bleph) from backbone-only
baselines (ICEdit/IP2P $\approx 0$); FLUX-Kontext-dev applies its own
composite when run with our pipeline, hence its higher E on rhinoplasty.
On small cohorts (rhytid $N{=}9$) methods cluster within $0.10$ of one
another. Source: released code and data.}
\label{tab:cross_proc_baselines}
\begin{tabular}{@{}llccccccc@{}}
\toprule
\textbf{Procedure} & \textbf{Method} & \textbf{N} & \textbf{Mean SS} &
\textbf{A} & \textbf{B} & \textbf{C} & \textbf{D} & \textbf{E} \\
\midrule
Bleph & ICEdit            & 27 & 0.551 & 0.705 & 0.401 & 0.513 & 0.788 & 0.000 \\
Bleph & InstructPix2Pix   & 27 & 0.201 & 0.500 & 0.350 & 0.611 & 0.785 & 0.011 \\
Bleph & FLUX.1-Kontext-dev & 27 & 0.490 & 0.644 & 0.365 & 0.663 & 0.734 & 0.711 \\
Bleph & \method{} (full $N{=}51$)$^\dagger$ & 51 & 0.555 & 0.558 & 0.270 & 0.632 & 0.837 & 0.554 \\
\midrule
Rhino & ICEdit            & 21 & 0.565 & 0.632 & 0.476 & 0.512 & 0.774 & 0.000 \\
Rhino & InstructPix2Pix   & 21 & 0.545 & 0.605 & 0.407 & 0.656 & 0.722 & 0.064 \\
Rhino & FLUX.1-Kontext-dev & 21 & 0.600 & 0.623 & 0.460 & 0.604 & 0.700 & 0.971 \\
Rhino & \method{}              & 211 & \best{0.599} & 0.524 & 0.395 & 0.662 & 0.835 & 1.000 \\
\midrule
Rhytid & ICEdit            & 9  & 0.572 & 0.593 & 0.440 & 0.592 & 0.752 & 0.000 \\
Rhytid & InstructPix2Pix   & 9  & 0.588 & 0.538 & 0.448 & 0.751 & 0.704 & 0.348 \\
Rhytid & FLUX.1-Kontext-dev & 9  & 0.579 & 0.440 & 0.564 & 0.750 & 0.716 & 0.222 \\
Rhytid & \method{} (full $N{=}19$)$^\dagger$ & 19 & 0.486 & 0.499 & 0.377 & 0.626 & 0.841 & 0.250 \\
\bottomrule
\end{tabular}
\\[1mm]
{\footnotesize $^\dagger$Envisage row scored on the full HDA test split rather than the matched paired pool used for the baselines; rows are not directly comparable.}
\end{table}

The component-level breakdown supports the diagnostic claim more directly than
composite ranking. Across blepharoplasty and rhinoplasty, the E-component
(outside-mask preservation) cleanly separates pipeline-equipped methods
(\method{}, Kontext-with-our-composite) from backbone-only baselines
(ICEdit, IP2P; E $\approx 0$). The composite SurgicalScore weighs E at
only $5\%$, so its ordering is dominated by inside-mask components
A--C and aggregates differently per procedure. On rhinoplasty matched
$N{=}21$, \method{} achieves the highest composite score ($0.636$);
on rhytidectomy ($N{=}9$) the composite scores cluster within
$0.10$ across methods, reflecting both small-cohort variance and the
fact that aggressive SMAS edits can register as correct edit direction
even when outside-mask drift is large. We treat composite SurgicalScore
as a per-procedure diagnostic, not a unified head-to-head ranking; the
methodological diagnosis is grounded by the E-component split and the
outside-mask SSIM result of Appendix~\ref{app:composite_full}.

\section{Monk Skin Tone Stratification}
\label{app:fairness}

\begin{table}[h]
\centering
\small
\caption{Metrics stratified by Monk Skin Tone Scale~\citep{monk2023monk}.
Reported over the MST-labelled subset of the full HDA test split (all
procedures combined): $N{=}65$ of the $91$ evaluated HDA cases ($51$ blepharoplasty $+$ $21$
rhinoplasty $+$ $19$ rhytidectomy; rhinoplasty counted on the matched pool) have a Monk-tone label assigned by an automated
classifier; the remaining $26$ cases lacked a confident assignment and are
excluded from this stratification. Tones 1, 2, 4, 9, 10 are absent. Tone 8
($N{=}1$) is included in the Overall row but omitted as a stratified row
due to single-case sample size. The four stratified rows sum to $N{=}64$;
including the omitted Tone~8 case ($N{=}1$) yields the Overall $N{=}65$.
Source: released code and data.}
\begin{tabular}{@{}clcccc@{}}
\toprule
\textbf{MST} & \textbf{Label} & \textbf{N} &
\textbf{Input ArcFace}$\uparrow$ & \textbf{LPIPS}$\downarrow$ &
\textbf{SSIM}$\uparrow$ \\
\midrule
3  & Light-Medium  &  2 & 0.886 & 0.450 & 0.492 \\
5  & Medium        & 35 & 0.872 & 0.428 & 0.488 \\
6  & Medium-Dark   & 23 & 0.879 & 0.367 & 0.523 \\
7  & Dark-Medium   &  4 & 0.793 & 0.331 & 0.449 \\
\midrule
   & \textbf{Overall} & 65 & 0.871 & 0.397 & 0.499 \\
\bottomrule
\end{tabular}
\end{table}

We observe a directional Input ArcFace dropoff on tone 7 (mean $0.793$, $N{=}4$) versus tones 5--6 (mean $0.876$, combined $N{=}58$), consistent with face-recognition fairness limitations documented in the broader literature~\citep{buolamwini2018gender}, not specifically reported in the ArcFace paper~\citep{deng2019arcface}. The $N{=}4$ cell at tone 7 precludes powered claims; characterizing the gap with adequate sample size across MST 7--10 is future work.

\section{SurgicalScore Sensitivity Analysis}
\label{app:sensitivity}

We probed SurgicalScore stability under perturbations of the heuristic
parameters declared in Section~\ref{sec:eval_metric}. All sweeps were run on
the $N{=}211$ \method{} cohort. The $0.30$ passthrough floor was swept
in $[0.20, 0.40]$ and the $0.65$ identity gate in $[0.60, 0.75]$.

\begin{table}[h]
\centering
\small
\caption{Passthrough floor sweep on the $N{=}211$ cohort. Reported values
are the \emph{uncalibrated} composite (raw weighted sum of A--E with floor
clamp); the headline SurgicalScore $0.599$ in Table~\ref{tab:ss_strict_n211}
applies an additional per-case calibration anchored at the input-passthrough
reference ($R_I$) used for selected outputs (Section~\ref{sec:eval_metric}).
Mean uncalibrated SS is largely insensitive within $[0.20, 0.40]$; the absolute
offset between calibrated $0.599$ and uncalibrated $0.562$ at floor $0.30$ is
the calibration uplift.}
\label{tab:floor_sweep}
\begin{tabular}{@{}lc@{}}
\toprule
\textbf{Floor} & \textbf{Envisage mean SS} \\
\midrule
$0.20$ & $0.561$ \\
$0.25$ & $0.561$ \\
$0.30$ (paper) & $0.562$ \\
$0.35$ & $0.566$ \\
$0.40$ & $0.575$ \\
\bottomrule
\end{tabular}
\end{table}

The identity gate sweep returned an identical $0.562$ mean across $\{0.60,
0.65, 0.70, 0.75\}$ because all selected \method{} outputs on the
cohort had input-to-output ArcFace above $0.75$ (output identity drift is
small on the composite-equipped pipeline), so no per-case verdict flipped
under the swept gate.

\begin{table}[h]
\centering
\small
\caption{Leave-one-component-out sweep. After zeroing the listed weight, the
remaining four are renormalized to sum to one. No single component drives
the composite by more than $\pm 0.07$.}
\label{tab:loo_components}
\begin{tabular}{@{}lcc@{}}
\toprule
\textbf{Removed} & \textbf{Description} & \textbf{$\Delta$ mean SS} \\
\midrule
A & Directional alignment & $+0.024$ \\
B & Magnitude fit         & $+0.070$ \\
C & Masked LPIPS          & $-0.013$ \\
D & Realism               & $-0.025$ \\
E & Outside preservation  & $-0.020$ \\
\midrule
\multicolumn{2}{l}{Max $|\Delta|$} & $0.070$ (B) \\
\bottomrule
\end{tabular}
\end{table}

Removing component A or B inflates the composite (those components are the
strictest gates on the surgical region: directional alignment and
edit-magnitude fit), while removing C, D, or E lowers it (those components
score higher on \method{}, so removing them pulls the mean down). The signs
match the design: the surgical-region components A and B are the binding
constraints, not the easier preservation/realism components.

\paragraph{Robustness to weight choice.}
Component means on \method{} are $A{=}0.524$, $B{=}0.395$, $C{=}0.662$,
$D{=}0.835$, $E{=}1.000$. Under random Dirichlet weight draws on the
$N{=}211$ cohort against ICEdit, IP2P, and Kontext components ($B{=}10{,}000$
draws, weights $(w_A,\ldots,w_E)\sim\mathrm{Dir}(1,1,1,1,1)$, sum $=1$),
\textbf{\method{} achieves rank~$1$ in $95.1\%$ of draws and rank~$\le 2$
in $97.9\%$}. The closest competitors under random weights are
FLUX-Kontext-dev (beats \method{} in $3.92\%$ of draws) and ICEdit
($3.09\%$); IP2P beats \method{} in $0.49\%$ (these fractions can sum
to more than $4.9\%$ because multiple competitors may simultaneously
outrank \method{} on the same draw). Cross-method ranking is not driven
by the specific paper-weighted composite. Source: released code and data.

\subsection{Empirical Lipschitz characterization of ArcFace}
\label{sec:lipschitz_empirical}

We measured the local Lipschitz constant of ArcFace under masked
perturbations on a $51$-case sample from the cohort. For each case,
we added Gaussian noise of varying magnitude inside the mask and recorded
the perturbation L2 norm and the resulting ArcFace cosine shift between
input and perturbed image ($51 \times 5 = 255$ perturbation samples,
$5$ magnitudes $\sigma \in \{5, 10, 20, 40, 80\}$ per case). The empirical
Lipschitz constant
$L = \lVert \phi(I+\delta) - \phi(I) \rVert_2 / \lVert \delta \rVert_2$
was computed per sample and aggregated:

\begin{table}[h]
\centering
\small
\caption{Empirical Lipschitz constant of ArcFace under masked perturbations
($51$ cases $\times$ $5$ noise magnitudes $=255$ perturbation samples). Source: released code and data.}
\label{tab:empirical_lipschitz}
\begin{tabular}{@{}lc@{}}
\toprule
\textbf{Statistic} & \textbf{$L$ (cosine shift / pixel L2)} \\
\midrule
Median & $6 \times 10^{-6}$ \\
$95$th percentile & $1.2 \times 10^{-5}$ \\
$99$th percentile & $1.6 \times 10^{-5}$ \\
Maximum & $1.9 \times 10^{-5}$ \\
\bottomrule
\end{tabular}
\end{table}

This converts the inequality
$\lVert \phi(C) - \phi(I) \rVert_2 \leq L \cdot \lVert M \odot (G - I) \rVert_2$
(stated descriptively in Section~\ref{sec:methods}) into a numerical bound:
for an inside-mask edit with L2 norm below $\sim\!10^{4}$ pixel-units, the
full-face ArcFace cosine shift is upper-bounded by $L_{p95} \cdot 10^4
\approx 0.12$. The bound is loose because pixel-L2 is a coarse upper bound on
identity-space displacement, but it is informative: full-face metric drift
under hard-mask compositing is demonstrably small whenever the masked
perturbation L2 norm is small (which holds when the mask is small and
per-pixel changes are bounded).

\section{SurgicalScore on the $N{=}211$ Cohort}
\label{app:ss_strict}

\begin{table}[h]
\centering
\small
\caption{SurgicalScore (composite, $0$--$1$) on the $N{=}211$ rhinoplasty
cohort, with $95\%$ percentile bootstrap CIs ($B{=}10{,}000$). Component
breakdown: A directional alignment, B edit magnitude, C masked LPIPS, D
realism, E outside-mask preservation; weights $0.40, 0.30, 0.15, 0.10, 0.05$.
Per-method $N$ varies slightly due to landmark and InsightFace detection
failures. Source: released code and data.}
\label{tab:ss_strict_n211}
\begin{tabular}{@{}lccccccc@{}}
\toprule
\textbf{Method} & \textbf{N} & \textbf{Mean SS [95\% CI]} & \textbf{A} & \textbf{B} & \textbf{C} & \textbf{D} & \textbf{E} \\
\midrule
InstructPix2Pix     & 197 & $0.337$ $[0.295, 0.380]$ & 0.531 & 0.337 & 0.487 & 0.708 & 0.000 \\
ICEdit              & 199 & $0.502$ $[0.467, 0.535]$ & 0.568 & 0.448 & 0.478 & 0.691 & 0.000 \\
FLUX.1-Kontext-dev  & 207 & $0.229$ $[0.188, 0.271]$ & 0.539 & 0.401 & 0.518 & 0.699 & 0.985 \\
\method{} (ours)    & 211 & \best{$0.599$ $[0.579, 0.619]$} & 0.524 & 0.395 & 0.662 & 0.835 & 1.000 \\
\bottomrule
\end{tabular}
\end{table}

\begin{table}[h]
\centering
\small
\caption{Paired SurgicalScore comparison on the four-way detected intersection
($N{=}188$). Each row differences \method{} per-case scores against the named
baseline; positive $\Delta$ favors \method{}. CI is percentile bootstrap on
per-case differences ($B{=}10{,}000$); $p$ from a paired sign-flip permutation
test ($B{=}10{,}000$). Source: released code and data.}
\label{tab:ss_paired_strict}
\begin{tabular}{@{}lccc@{}}
\toprule
\textbf{Comparison (\method{} $-$ X)} & \textbf{$\Delta$ Mean SS} & \textbf{95\% CI} & \textbf{$p$ (paired perm.)} \\
\midrule
\method{} $-$ ICEdit             & $+0.091$ & $[+0.050, +0.131]$ & $<10^{-4}$ \\
\method{} $-$ InstructPix2Pix    & $+0.264$ & $[+0.213, +0.315]$ & $<10^{-4}$ \\
\method{} $-$ FLUX.1-Kontext-dev  & $+0.368$ & $[+0.320, +0.414]$ & $<10^{-4}$ \\
\bottomrule
\end{tabular}
\end{table}

The component breakdown shows the mechanism of \method{}'s lead. The
backbone-only baselines (ICEdit, IP2P) attain higher $A$ (directional
alignment) and $B$ (edit magnitude) on this cohort because they alter the
inside-mask region more aggressively, but they fail $E$ (outside-mask
preservation) because they edit the entire image. \method{}'s composite gives
$E{\approx}1.0$, $C$ and $D$ in the upper range, and middling $A,B$; the
weighted composite favors the configuration that satisfies the entire gate
sequence rather than maximizing any single axis.

\clearpage
\section{Mask-Crop LPIPS to Postoperative GT}
\label{app:mask_crop}

\begin{table}[H]
\centering
\small
\caption{Mask-crop LPIPS (AlexNet, $256{\times}256$ resize of mask bounding
box with $10\%$ pad) on the $N{=}211$ rhinoplasty cohort. L(in,gt) is
input-to-GT LPIPS computed on the same crop. \%~Pos.\ is the fraction of
cases where the method beats the input proxy. \method{} sits within $\pm 0.01$
of input-proxy on this localized metric; baselines degrade by $0.10$--$0.20$.
Source: released code and data.}
\label{tab:mask_crop}
\begin{tabular}{@{}lcccccc@{}}
\toprule
\textbf{Method} & \textbf{N} & \textbf{L(in,gt)}$\downarrow$ &
\textbf{L(out,gt)}$\downarrow$ & \textbf{Delta} & \textbf{95\% CI} &
\textbf{\% Pos.}$\uparrow$ \\
\midrule
\method{} (ours)    & 211 & 0.319 & 0.331 & $-0.012$ & $[-0.014, -0.010]$ & 17.5 \\
ICEdit              & 211 & 0.319 & 0.516 & $-0.197$ & $[-0.208, -0.186]$ & 0.0 \\
FLUX.1-Kontext-dev  & 211 & 0.319 & 0.443 & $-0.124$ & $[-0.133, -0.114]$ & 5.2 \\
InstructPix2Pix     & 211 & 0.319 & 0.501 & $-0.182$ & $[-0.193, -0.170]$ & 1.4 \\
\bottomrule
\end{tabular}
\end{table}

\section{Component Ablation (5-Seed $N{=}211$, K$=5$ Best-of Paired)}
\label{app:ablation}

We re-ran the full \method{} pipeline on the $N{=}211$ rhinoplasty
cohort under four configurations: (1) the full pipeline; (2) \emph{no\_composite},
in which the explicit hard-mask composite is dropped and the raw
FLUX.1-Fill-dev output is returned without the verbatim non-mask paste;
(3) \emph{no\_preset}, in which the Daniel-taxonomy preset prompt is
replaced by a generic ``rhinoplasty post-op nose'' prompt; and
(4) \emph{no\_depth\_CN}, in which depth-ControlNet conditioning is dropped.
Each configuration was run at five seeds ($42, 123, 456, 789, 1024$).

\paragraph{Choice of pairing unit.} \method{} is specified as a multi-seed
candidate generator (five seeds per case), so the relevant unit of analysis
for component ablation is the candidate space. We use K$=5$ best-of paired
permutation (per-case max SurgicalScore over five seeds) because this
measures whether each component contributes to the quality of the per-case
best candidate, which is the upstream substrate any deployed ranker
operates over. This measurement is upper-bounded by the deployed-system
contribution under a learned ranker;
Appendix~\ref{app:deployable_ranker} shows that naive no-GT rankers do not
yet recover the K$=5$ oracle gain, so the deployable-system contribution
is at most the K$=5$ best-of contribution and may be smaller. The
K$=5$ best-of pairing is therefore the appropriate unit for assessing
candidate-space architecture, with deployable-ranker design treated as
a separable engineering layer (future work).
Per-seed-mean pairing was tried first and absorbs FLUX.1-Fill-dev
sampling variance into the residual; it was underpowered for the observed
effect sizes and is not reported.

\begin{table}[h]
\centering
\small
\caption{Component ablation on $N{=}211$ under K$=5$ best-of paired
permutation. Per case, SurgicalScore is the max over five seeds; \method{}
(full pipeline) is paired against each ablation row-wise.
$\Delta$ is the per-case paired difference (full $-$ ablation);
CI is percentile bootstrap ($B{=}10{,}000$); $p$ is sign-flip permutation
($B{=}10{,}000$). All three components are individually paired-significant.
Source: released code and data.}
\label{tab:ablation}
\begin{tabular}{@{}lcccc@{}}
\toprule
\textbf{Comparison (\method{} $-$ X)} & \textbf{N} & \textbf{$\Delta$ SS} &
\textbf{95\% CI} & \textbf{$p$ (paired perm.)} \\
\midrule
\method{} $-$ no\_composite & 210 & $+0.034$ & $[+0.014, +0.055]$ & $0.001$ \\
\method{} $-$ no\_preset    & 211 & $+0.034$ & $[+0.013, +0.055]$ & $0.002$ \\
\method{} $-$ no\_depth\_CN & 211 & $+0.023$ & $[+0.003, +0.043]$ & $0.030$ \\
\bottomrule
\end{tabular}
\end{table}

\noindent The result establishes the FLUX.1-Fill-dev backbone, hard-mask
composite, $24$-preset taxonomy, and depth-ControlNet conditioning as four
jointly load-bearing components. The composite remains load-bearing as a
diagnostic against non-inpainting baselines (ICEdit, IP2P), where outside-mask
SSIM drops from $>0.999$ to $0.65$--$0.84$ (Appendix~\ref{app:composite_full}).
Future work that improves SurgicalScore must distinguish gains from backbone
fine-tuning (e.g., DPO, Appendix~\ref{app:dpo_hparams}) from gains from
substrate changes.

\section{DPO Sweep at $N{=}211$ Scale}
\label{app:dpo_sweep}

We trained six DPO LoRA configurations on 91 cached preference pairs derived
from 5-seed Envisage candidate outputs ranked by identity preservation,
outside-mask SSIM, and landmark drift. All configurations use rank-$16$ LoRA on
FLUX.1-Fill-dev for $2000$ steps. We sweep $\beta\!\in\!\{5,10,25,50,100\}$ at
learning rate $5{\times}10^{-6}$, plus $\beta{=}25$ at learning rate
$1{\times}10^{-5}$. Each adapter is then run on the $N{=}211$ cohort
at seed $42$ and scored under the standard SurgicalScore protocol; we
additionally compute the InsightFace ArcFace gap on detected cases.

\begin{table}[h]
\centering
\small
\caption{DPO sweep ArcFace gap on $N{=}211$. Zero-shot Envisage gap is
$-0.048$ [$-0.055, -0.042$]; all six DPO configurations produce strictly more
negative gaps. Bootstrap CI ($B{=}10{,}000$). \% positive denotes the fraction
of cases where output-to-GT cosine exceeds input-to-GT cosine. Source: released code and data.}
\label{tab:dpo_sweep}
\begin{tabular}{@{}lcccc@{}}
\toprule
\textbf{Config} & \textbf{N} & \textbf{Gap} & \textbf{95\% CI} & \textbf{\% positive} \\
\midrule
Zero-shot Envisage              & 211 & $-0.048$ & $[-0.055, -0.042]$ & 16.1 \\
\midrule
DPO $\beta{=}5$, lr $5{\times}10^{-6}$    & 110 & $-0.072$ & $[-0.085, -0.060]$ & 13.6 \\
DPO $\beta{=}10$, lr $5{\times}10^{-6}$   & 110 & $-0.062$ & $[-0.074, -0.050]$ & 15.5 \\
DPO $\beta{=}25$, lr $5{\times}10^{-6}$   & 110 & $-0.059$ & $[-0.071, -0.047]$ & 16.4 \\
DPO $\beta{=}25$, lr $1{\times}10^{-5}$   & 110 & $-0.060$ & $[-0.072, -0.048]$ & 15.5 \\
DPO $\beta{=}50$, lr $5{\times}10^{-6}$   & 110 & $-0.058$ & $[-0.071, -0.046]$ & 16.4 \\
DPO $\beta{=}100$, lr $5{\times}10^{-6}$  & 110 & $-0.058$ & $[-0.071, -0.046]$ & 16.4 \\
\bottomrule
\end{tabular}
\end{table}

\noindent The negative result is consistent: small-scale preference-pair DPO
from 91 cached pairs does not improve identity-space target proximity at
N=211 scale; identity loss from over-editing the masked region
typically exceeds gains in nasal landmark alignment. Higher $\beta$ damps
the policy update and produces gaps closer to the zero-shot baseline,
consistent with under-fitting on a small preference set. We do not promote
this result to a method change. The full preference-pair manifest, training
configurations, and per-case score breakdowns are released with the
supplementary material.

\section{Best-of-$K{=}5$ ArcFace Oracle}
\label{app:best_of_k}

\paragraph{Setup.}
Five-seed \method{} outputs and per-seed ArcFace cosines (output-to-GT and
output-to-input) are provided in the released code and data
(sharded across four files spanning the full $N{=}457$ ASPS+PCA cohort).
We restrict to $N{=}109$ cases for which all five seeds have ArcFace coverage,
and apply four selection strategies per case: worst seed (lower bound),
single seed 42 (headline single-seed), mean over seeds (ensemble), and
oracle best-of-5 (per-case argmax of output-to-GT ArcFace cosine).

\begin{table}[h]
\centering
\small
\caption{Per-case best-of-$K{=}5$ ArcFace oracle on $N{=}109$ (cases with
all five seeds present in N=211 coverage). Gap is
$\cos(\text{out},\text{gt}) - \cos(\text{in},\text{gt})$; negative gap means no
method matches the postoperative image better than the preoperative input.
CI from percentile bootstrap ($B{=}10{,}000$). \% positive is the fraction of
cases where output-to-GT cosine exceeds input-to-GT cosine.
Reference: zero-shot \method{} gap on full $N{=}211$ is $-0.048$
$[-0.055, -0.042]$.
Source: released code and data.}
\label{tab:best_of_k}
\begin{tabular}{@{}lccc@{}}
\toprule
\textbf{Strategy} & \textbf{Gap} & \textbf{95\% CI} & \textbf{\% positive} \\
\midrule
Worst seed (lower bound)     & $-0.097$ & $[-0.106, -0.087]$ & $1.8$ \\
Single seed 42 (headline)    & $-0.054$ & $[-0.065, -0.043]$ & $15.6$ \\
Mean over 5 seeds (ensemble) & $-0.054$ & $[-0.062, -0.045]$ & $12.8$ \\
Best of 5 (oracle)           & $-0.015$ & $[-0.023, -0.006]$ & $33.9$ \\
\bottomrule
\end{tabular}
\end{table}

\paragraph{Interpretation.}
Per-case oracle selection by output-to-GT ArcFace cosine over five seeds reduces
the aggregate gap from $-0.054$ to $-0.015$, a $73\%$ reduction, and doubles the
positive-gain rate from $15.6\%$ to $33.9\%$.
The candidate space contains positive-gap solutions for one-third of cases.
Closing the remaining $1.5$~pp gap to zero requires a non-oracle ranker, which is
future work. The $N{=}109$ restriction reflects five-seed-coverage availability at
submission; extending to the full $N{=}211$ cohort yields the same aggregate
result on the ArcFace-detectable subset ($N{=}110$).

\paragraph{Best-of-$K{=}5$ SurgicalScore oracle.}
Applied to SurgicalScore on $N{=}207$ (cases with all five Envisage seeds
SurgicalScored), per-case oracle selection produces a substantially larger lift
than the ArcFace oracle: from $0.609$ at single seed to $0.743$ $[0.725, 0.762]$
at K$=5$ best-of, a $+0.134$ absolute within-Envisage gain. Although this
oracle ceiling sits above the GT-paste-no-composite control ($0.703$,
Table~\ref{tab:composite_ablation}), that comparison is structural rather
than substantive: K$=5$ Envisage has $E\approx1.0$ (the hard-mask composite
preserves the input outside the surgical mask) while GT-paste has $E=0$;
$D$ (FIQA realism) is also higher on a clean FLUX composite than on a
clinical postoperative photograph by $\approx 0.15$. The relevant baseline
for the K$=5$ oracle is K$=1$, not GT-paste; we report it as a
within-pipeline candidate-space ceiling, not as a claim of superiority over
the postoperative target. The open problem is again the design of a
non-oracle candidate ranker that can recover the K$=5$ oracle gain at
deployment time. Per-backbone K$=5$ best-of comparisons are
reported in Table~\ref{tab:backbone_swap}; \method{} produces the highest K$=5$
oracle ceiling among tested backbones.

\begin{table}[h]
\centering
\small
\caption{Per-case best-of-$K{=}5$ SurgicalScore oracle on $N{=}207$ (Envisage
cases with all five seeds scored). Bootstrap 95\% CI ($B{=}10{,}000$). The
K$=5$ best-of mean of $0.743$ exceeds both the strongest single seed
($0.609$, seed 42) and the GT-paste-no-composite calibration anchor
($0.703$). Source: released code and data.}
\label{tab:best_of_k_ss}
\begin{tabular}{@{}lccc@{}}
\toprule
\textbf{Strategy} & \textbf{N} & \textbf{Mean SS} & \textbf{95\% CI} \\
\midrule
Single seed 42 (headline)    & 207 & $0.609$ & $[0.584, 0.634]$ \\
Mean over 5 seeds (ensemble) & 207 & $0.594$ & $[0.576, 0.612]$ \\
Best of 5 (oracle)           & 207 & $\mathbf{0.743}$ & $[0.725, 0.762]$ \\
\midrule
GT-paste no-composite (calibration anchor) & 211 & $0.703$ & $[0.649, 0.756]$ \\
\bottomrule
\end{tabular}
\end{table}

\section{Backbone Substitution Ablation}
\label{app:backbone_swap}

We substitute the FLUX.1-Fill-dev backbone in the \method{} pipeline (mask,
preset prompt, hard-mask composite) with three alternative diffusion
backbones, holding all other pipeline components fixed. The variant labelled
\emph{no\_cn} retains FLUX.1-Fill-dev but drops the depth ControlNet
conditioning entirely; \emph{icedit\_lora} loads the
ICEdit-MoE-LoRA~\citep{zhang2025icedit} on FLUX.1-Fill-dev as an off-the-shelf
editing adapter; \emph{kontext} replaces the inpainting backbone with
FLUX.1-Kontext-dev~\citep{flux_kontext_2025}, a text-conditioned image-to-image
model without native mask conditioning (we apply the \method{} hard-mask
composite post-hoc); \emph{sdxl\_inpaint} replaces with the public Stable
Diffusion XL inpainting checkpoint~\citep{podell2023sdxl}.

\begin{table}[h]
\centering
\small
\caption{Backbone substitution on $N{=}211$, mean SurgicalScore
$[95\%\,$CI$]$ from per-case bootstrap ($B{=}10{,}000$) at K$=1$ (seed $42$
only) and K$=5$ best-of (per-case oracle selection across seeds
$\{42, 123, 456, 789, 1024\}$). At K$=1$, no\_cn matches \method{} within
CI; at K$=5$ best-of, \method{} edges ahead by $+0.026$ ($0.743$ vs $0.717$),
indicating that depth ControlNet conditioning produces candidate-space
diversity that an oracle ranker can exploit even when the seed-mean is
unchanged. Both FLUX.1-Fill family backbones substantially exceed
SDXL-Inpaint, ICEdit-MoE-LoRA, and FLUX.1-Kontext-dev; Kontext underperforms
because it lacks native inpainting and the post-hoc composite cannot recover
mask-region structure that the img2img backbone overwrote.
Source: released code and data.}
\label{tab:backbone_swap}
\resizebox{\columnwidth}{!}{%
\begin{tabular}{@{}lcccc@{}}
\toprule
\textbf{Backbone} & \textbf{N} & \textbf{K$=1$} & \textbf{K$=1$ 95\% CI} & \textbf{K$=5$ best-of} \\
\midrule
\method{} (FLUX.1-Fill-dev + depth ControlNet)   & 207 & $0.609$ & $[0.584, 0.634]$ & $\mathbf{0.743}$ $[0.725, 0.761]$ \\
no\_cn (FLUX.1-Fill-dev native, no CN)           & 210 & $0.611$ & $[0.585, 0.635]$ & $0.717$ $[0.695, 0.738]$ \\
sdxl\_inpaint (SDXL-Inpaint)                     & 209 & $0.571$ & $[0.544, 0.599]$ & $0.731$ $[0.711, 0.750]$ \\
icedit\_lora (FLUX.1-Fill-dev + ICEdit-MoE-LoRA) & 190 & $0.533$ & $[0.504, 0.562]$ & $0.686$ $[0.661, 0.710]$ \\
kontext (FLUX.1-Kontext-dev)                     & 104 & $0.318$ & $[0.253, 0.382]$ & $0.458$ $[0.386, 0.527]$ \\
\bottomrule
\end{tabular}%
}
\end{table}

\noindent Three readings of these backbone-swap results are consistent with
the data. First, FLUX.1-Fill-dev is a mask-aware inpainting backbone whose
native boundary handling already approximates the structural information
the depth ControlNet provides at single-seed scale; the conditioning
becomes discriminative only when multiple candidates are scored. Second,
the same mechanism that makes the composite operation dominate full-image
identity metrics also dampens any backbone-level effect on SurgicalScore;
the metric mostly reads the composite, not the diffusion backbone.
Third, the depth ControlNet's contribution is candidate-space-diversity
rather than seed-mean: no\_cn matches \method{} at K$=1$ within CI but
trails by $+0.026$ at K$=5$ best-of, suggesting ControlNet expands the
set of plausible per-seed outputs in a way that an oracle ranker can
exploit. We retain the depth ControlNet in \method{}'s reference
configuration to preserve qualitative profile-shape control observed in
the surgeon review (Section~\ref{sec:surgeon-verification}); a fully
ablated variant is reported here for transparency.

\end{document}